\documentclass[pre,aps,floatfix,twocolumn,showpacs,superscriptaddress]{revtex4-2} 
\usepackage{color,soul}
\usepackage{graphicx}
\usepackage{amsmath, amsthm, amssymb,mathtools}
\usepackage{amsthm}
\usepackage{multirow}
\usepackage[normalem]{ulem}
\usepackage{soul}
\usepackage{hyperref}
\newcommand{\be}{\begin{equation}}
\newcommand{\ee}{\end{equation}}
\newcommand{\bea}{\begin{eqnarray}}
\newcommand{\eea}{\end{eqnarray}}
\newcommand{\eref}[1]{Eq.~(\ref{#1})}%
\makeatletter

\newcommand{\Rmnum}[1]{\expandafter\@slowromancap\romannumeral #1@}
\makeatother\usepackage{array, makecell} 

\begin{document}

\title{Mpemba effect on non-equilibrium active Markov chains}

\author{Apurba Biswas}
\email{apurba.biswas@u-bordeaux.fr}
\affiliation{Laboratoire Ondes et Matière d’Aquitaine, CNRS, UMR 5798, Université de Bordeaux, F-33400 Talence, France}
\author{Arnab Pal} 
\email{arnabpal@imsc.res.in}
\affiliation{The Institute of Mathematical Sciences, C.I.T. Campus, Taramani, Chennai 600113, India}
\affiliation{Homi Bhabha National Institute, Training School Complex, Anushakti Nagar, Mumbai 400094, India}


\begin{abstract}
We study the Mpemba effect on a non-equilibrium Markov chain that imitates the active motion of random walkers in a discrete energy landscape. The broken detailed balance, rendered by the activity,  gives rise to a unique anomalous relaxation in the system which is distinctly different than the typical equilibrium systems. We observe that the activity can both suppress or induce the Mpemba effect. Furthermore, we report an oscillatory Mpemba effect where the relaxation trajectories, emanating from the \textit{hot} and \textit{cold} initial conditions, cross each other multiple times and this occurs due to the emergence of complex eigenvalues in the relaxation spectrum due to the activity. Our work reveals a possible pathway for studying the Mpemba effect in active living systems where broken detailed balance is crucial to achieve many biological functions.
\end{abstract}

\maketitle

\section{Introduction}
The Mpemba effect refers to the anomalous relaxation phenomena where an initially hotter system equilibrates faster than an initially cooler system, when both systems are quenched to the same low temperature~\cite{Mpemba_1969}. Although the phenomena was first observed in the case of water namely, hot water freezing faster~\cite{Mpemba_1969, gijon2019paths} than an equal volume of  warm water exposed to the same condition, it is now a well-observed phenomena that has been probed through both theoretical and experimental means across various systems~\cite{chaddah2010overtaking,kumar2020exponentially,kumar2021anomalous, Walker_2021,Busiello_2021,lapolla2020faster,walker2022mpemba,degunther2022anomalous,PhysRevLett.124.060602,Klich-2019,das2021should,teza2021relaxation,Lu-raz:2017,PhysRevResearch.3.043160,schwarzendahl2021anomalous,SpinGlassMpemba,PhysRevE.104.064127,PhysRevLett.127.060401,chatterjee2023quantum,shapira2024mpemba,holtzman2022landau,Lasanta-mpemba-1-2017,mompo2020memory,PhysRevE.102.012906,biswas2021mpemba,biswas2022mpemba,megias2022mpemba,biswas2023measure,biswas2023mpemba,PhysRevE.108.024131,PhysRevLett.132.117102}. A particular frontier that remains notably unexplored is the effect of \textit{activity} in the anomalous relaxation dynamics (besides the work by Schwarzendahl and Löwen on active colloids \cite{schwarzendahl2021anomalous}). Activity is a hallmark feature of non-equilibrium systems  
which unlike their equilibrium counterpart violate detailed balance and this results in many interesting features such as non-Boltzmannian steady states, clustering, spontaneous aggregation, motility induced phase transition in active self-propelled systems, sense and adaption for high fidelity DNA transcription, structural organization in cells, folding of macromolecules \cite{gnesotto2018broken}. In this paper, we aim to understand how the broken detailed balance, specifically induced due to the \textit{activity}, can manifest in the anomalous relaxation dynamics such as the Mpemba effect.

In particular, we show via a rigorous analysis that an active system can display a rich behaviour in the relaxation dynamics such as the emergence of oscillations in the transient and the activity -induced or -suppressed Mpemba effect. The results are intriguing as the presence of such behaviours in the eigenspectrum or in the relaxation dynamics has no counterpart in regular equilibrium setups. Although a similar oscillatory behaviour in the relaxation has been reported recently in a two-level quantum system where such behaviour is attributed to the presence of exceptional points \cite{chatterjee2024multiple}, the presence of such behaviours in the classical models is rather unique which we explore herein. Finally, the emergence of non-equilibrium steady states and the oscillatory relaxations raises plausible questions
on the adaptable measures (pertaining only to the truncated eigenspectrum) that are often used to quantify the Mpemba effect in equilibrium systems~\cite{Klich-2019,Lu-raz:2017}.

\begin{figure}
\centering
\includegraphics[width=\columnwidth]{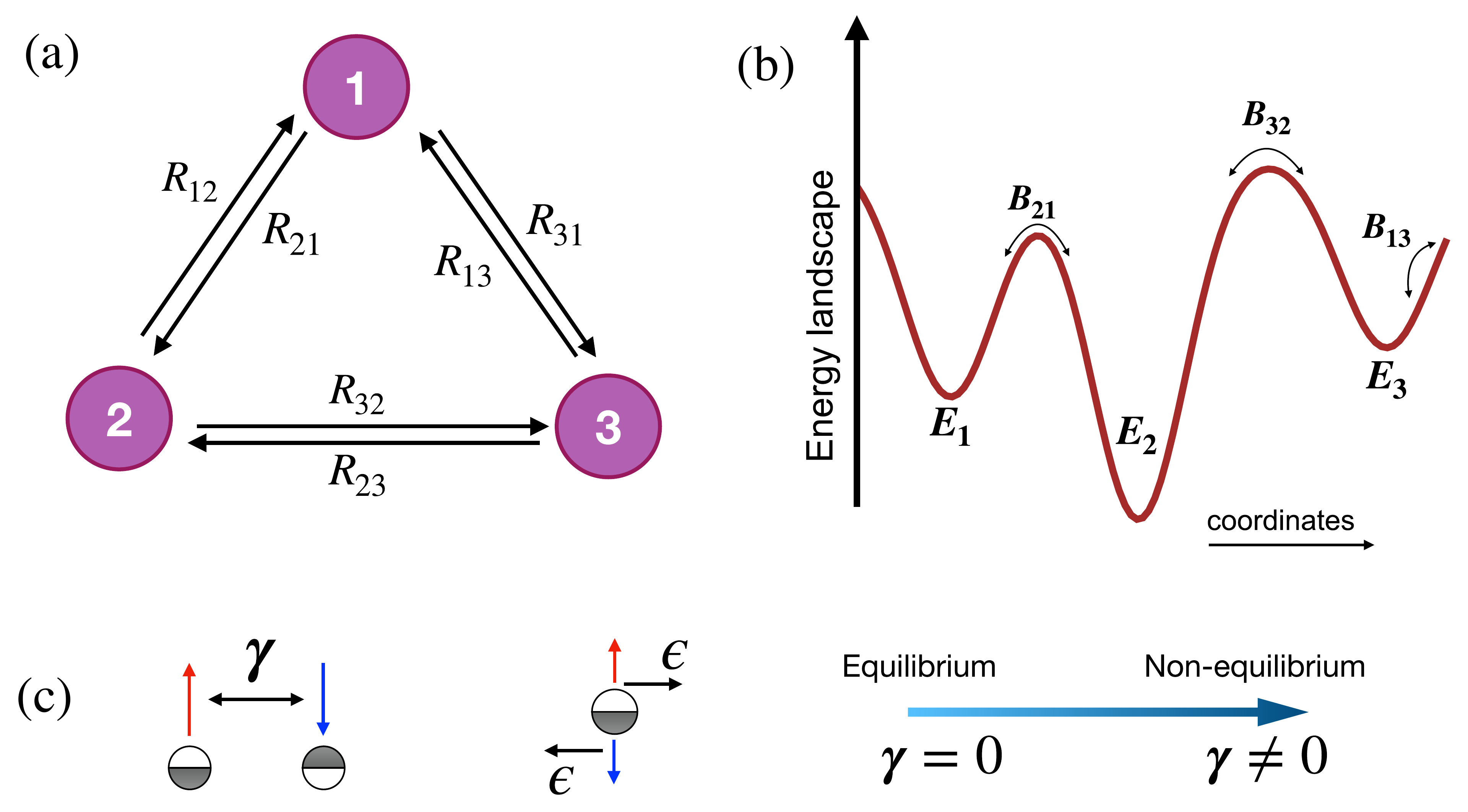} 
\caption{\label{3 state}Illustration of an active Markov chain model capturing the dynamics of a run and tumble motion of an active walker in a discrete 3-state energy landscape. Panel (a): The hopping rates $R_{ij}$ that satisfy detailed balance and in effect, characterize the rates of diffusion of an active walker among the energy states. Panel (b) depicts the corresponding energy landscape where $B_{ij}$'s are the energy barriers between states with energies $E_i$ and $E_j$. Panel (c): The activity of the walker is characterized by the rates $\gamma$ and $\epsilon$, where $\gamma$ is the flipping rate between the states $\uparrow$ and $\downarrow$ -- equivalent to the persistence between the two states, and $\epsilon$ is the state-dependent rate of unidirectional hopping that resembles the velocity of the active walker. In essence, $\gamma \neq 0$ is the non-equilibrium active component that drives the system away from  equilibrium.}
\end{figure}

The remainder of the paper is organized as follows. Section~\ref{sec:model} contains the model and its framework. The general definition of the Mpemba effect and the criteria for its occurrence are discussed in Sec.~\ref{sec:mpemba effect}. Next, we describe the effect of activity on the Mpemba effect along with qualitative reasoning in Sec.~\ref{sec:activity harnessed mpemba effect}.  Section~\ref{sec:oscillatory mpemba} describes the oscillatory Mpemba effect and the effect of activity on its phase diagram in terms of the model parameters, and a comparison between the oscillatory Mpemba effect and a damped harmonic oscillator is discussed in Sec.~\ref{sec:comparison with damped oscillator}. Section~\ref{sec:multiple crossing with real eigenspectrum} illustrates the possibility of having multiple crossings of trajectories even in the presence of real eigenspectrum and Sec.~\ref{Conclusion} contains the summary of results
and a discussion of their implications. 

\section{Model and framework}\label{sec:model}
Consider a setup of a discrete three state active Markov chain (Fig.~\ref{3 state}). The intrinsic hopping rate of a random walker from the 
state $j$ to $i$ is denoted by $R_{ij}$ and vice-versa. These rates satisfy detailed balance and are determined at a fixed bath temperature $T_b$ by the following relation 
\begin{align}
\label{DB}
\begin{split}
R_{ij}=e^{-\frac{B_{ij}-Ej}{k_B T_b}}, \text{ for } i\neq j,\\
R_{ji}=-\sum_{k\neq i}R_{ki}, \text{ for } i= j,
\end{split}
\end{align}
where $E_i$ are the energy levels of the three states, $B_{ij}$  are the barrier heights between any two states $i$ and $j$ and $k_B$ is Boltzmann constant. These rates usually characterize the rates of diffusion of the walker among the different energy states. To introduce the activity, we assume that there is an internal spin degree of freedom (denoted by $\uparrow$ and $\downarrow$) associated with each state. The spin can flip with a rate $\gamma$ which characterizes the persistence of the walker to be in that particular internal state. Depending on its current internal state ($\uparrow$ or $\downarrow$), the walker is biased to hop either clockwise, i.e., in the direction of $1\rightarrow 2\rightarrow 3\rightarrow 1$ for $\uparrow$ state or anti-clockwise for $\downarrow$ state at a rate $\epsilon$, thus characterizing the velocity of the active walker \cite{jose2022active}. 
Notably, our system stems from the active lattice gas models with Ising interaction that were proposed to describe flocking \cite{solon2013revisiting, solon2015flocking}.
The Markov chain kinetic model used here are also fundamental to prototypical unimolecular enzyme reaction cycles or cellular biochemical switches that can be out of chemical equilibrium due to chemical-potential difference and consequently, the rates will not be detailed balanced (see \cite{qian2007phosphorylation} for an extensive review on this topic). Note that the breakdown of detailed balance is explicit in our system as the diffusive rates are energy balanced while the rates for flipping and directional hopping are not.  In our analysis, we will be using the dimensionless variables. The time is measured in units of $\epsilon^{-1}$, the energy levels and barrier heights are measured in units of $k_B T_b$ but we set $k_B=1$ for simplicity. Any initial temperature $T$ is measured in units of the final bath temperature $T_b$, and the activity rates $\epsilon$ and $\gamma$ are measured in units of the intrinsic rate $R_{12}(T_b)$.

The central quantity required to understand the anomalous relaxation is the joint occupation probability $p_{i,\sigma}(t)$ which estimates the probability for the walker to be in the site  $i$ in $\sigma$ state, where $\sigma =\{ \uparrow, \downarrow \}$. The master equations for $p_{i,\sigma}(t)$ at time $t$ can be constructed from the knowledge of the hopping and active rates as
\begin{flalign}
\frac{d p_{1,\sigma}}{d t}=&(R_{12}\pm \epsilon) p_{2,\sigma} + (R_{13} \mp \epsilon) p_{3,\sigma}
-(R_{21}+R_{31}) p_{1,\sigma} \nonumber \\ &-\gamma p_{1,\sigma}+ \gamma p_{1,\bar{\sigma}},
\label{occu-1}
\end{flalign}
and similarly, for the sites $i=2,3$ (see Appendix~\ref{sec:solution} for details). Here, $\bar{\sigma}$ denotes the conjugate of $\sigma$. To solve for the occupation probability, it is useful to define the following quantities
\begin{align}
P_i=p_{i,\uparrow}+p_{i,\downarrow}, \nonumber\\
P^*_i=p_{i,\uparrow}-p_{i,\downarrow},
\label{occu-2}
\end{align}
where $P_i$ denotes the total probability of being at site $i$ summed over the internal spin states $\uparrow$ or $\downarrow$, and $P^*_i$ can be understood as the polarisation at site $i$ and its positive value denotes the preference of state $\uparrow$ over $\downarrow$, and vice-versa. Equation (\ref{occu-2}) can now be written in a concise matrix form
\begin{align}
\frac{d \boldsymbol{P} (T,t)}{dt}=\boldsymbol{W}(T_b)  \boldsymbol{P}(T,t), \label{matrix form}
\end{align}
where the vector $\boldsymbol{P}=(P_1, P_2, P_3, P^*_1, P^*_2, P^*_3)^{\mathbb{T}}$ denotes the instantaneous probability distributions of the system at time $t$ and $\boldsymbol{W}(T_b)$ is the transition matrix that is determined at the bath temperature $T_b$ [the form for $\boldsymbol{W}(T_b)$ is given in Eq.~(\ref{sm: W matrix})]. Here, the initial condition is chosen to be a steady state probability distribution $\boldsymbol{P}(T,t=0)$ characterized by a different bath temperature $T\neq T_b$ to which the system is initially prepared before being quenched to the final bath temperature $T_b$. If the transitions between microscopic states are pairwise-balanced [like in Eq. (\ref{DB})], the system reaches a thermodynamic equilibrium precluding
net flux among the states. On the other hand, if the transitions are not balanced, the system can exhibit flux loops, and moreover attains a non-equilibrium steady state.

\begin{figure}
\centering
\includegraphics[width=\columnwidth]{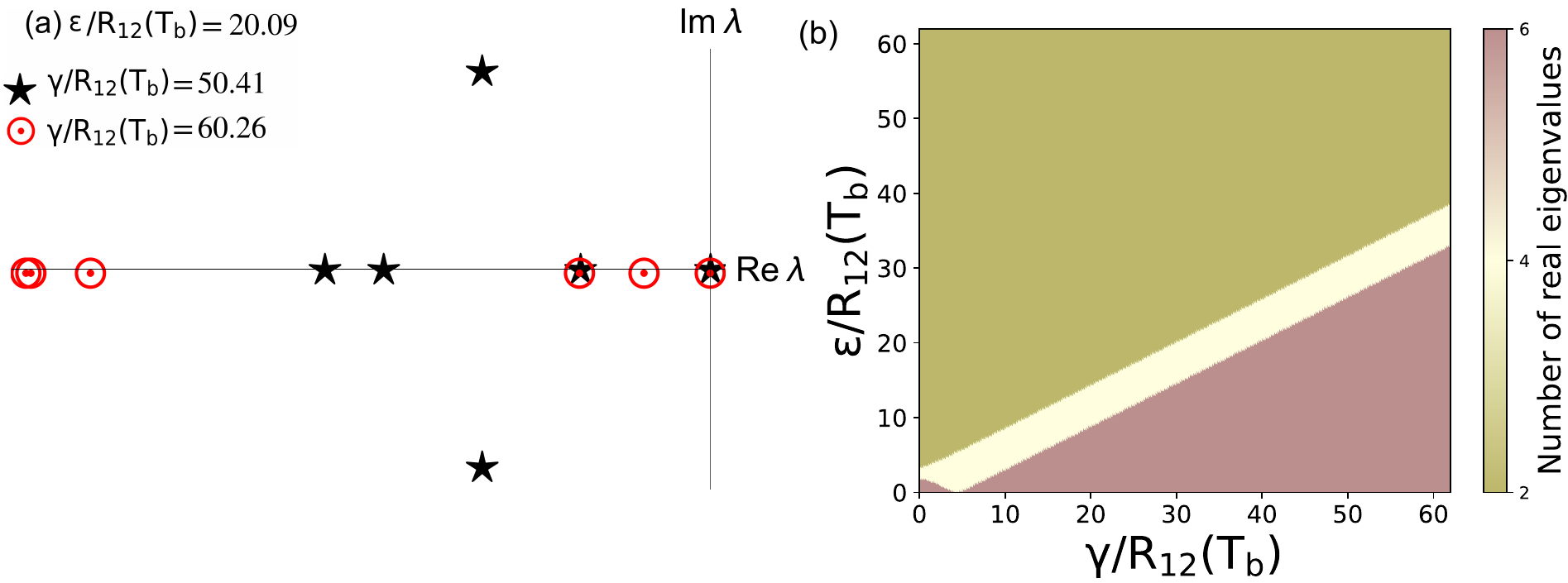} 
\caption{\label{eigenspectrum}Eigenvalues $\{ \lambda \}$-s of the transition matrix $\boldsymbol{W}$ of an active Markov chain (with broken detailed balance): Presence of both real and  complex eigenvalues is observed as we change the activity parameters. The transition from real to mixed eigenvalues as $\gamma$ is varied for fixed $\epsilon/R_{12}(T_b)=20.09$ [(in panel (a)]. 
Panel (b) explores the behavior in the number of real eigenvalues, from two (minimum) to six (maximum), as a function of $\epsilon$ and $\gamma$. The other relevant parameters are set as in Figs.~\ref{oscillatory mpemba diff measures} and \ref{phase_diagram}.
}
\end{figure}

Evidently, with the above-mentioned dynamics with induced activity (hence, broken detailed balance), the time evolution of the distribution $\boldsymbol{P}(T, t)$  attains a steady state distribution, $\text{lim}_{t\to \infty} \boldsymbol{P}(T,t)=\boldsymbol{\pi}(T_b)$ at large times, corresponding to the final quenched temperature of the bath $T_b$. During the relaxation process, the time evolution of the initial distribution $\boldsymbol{P}(T,t=0)$ is obtained by solving the time-dependent Eq.~(\ref{matrix form}) and is given by
\begin{flalign}
\boldsymbol{P}(T, t)=\boldsymbol{\pi}(T_b)+\sum^{6}_{i=2}a_i (T, T_b) \boldsymbol{v}_i e^{\lambda_i t}, \label{time evolution}
\end{flalign}
where $\lambda_i$ and $\boldsymbol{v}_i$ are the eigenvalues and right eigenvectors of the transition matrix $\boldsymbol{W}(T_b)$ respectively such that
\begin{equation}
\boldsymbol{W} (T_b) \boldsymbol{v}_i = \lambda_i \boldsymbol{v}_i.
\end{equation}
The eigenvalues follow the order $\lambda_1=0>\text{Re}\lambda_2\geq \text{Re}\lambda_3 \ldots,$ where $\lambda_1=0$ is the largest eigenvalue
that corresponds to the eigenvector $\boldsymbol{\pi}(T_b)$. In Eq.~(\ref{time evolution}), the coefficient $a_i(T,T_b)=\boldsymbol{v}'_i. \boldsymbol{X}^{-1} \boldsymbol{P}(T,t=0)/\boldsymbol{v}'_i.\boldsymbol{v}'_i$ is the overlap coefficient of the $i^{th}$ eigenvector $\boldsymbol{v}'_i$ with the initial distribution $\boldsymbol{P}(T, t=0)$. Here, $\boldsymbol{X}$ is formed from the eigenvectors $\boldsymbol{v}$ of $\boldsymbol{W}$ and the transformed eigenvectors $\boldsymbol{v}'$ are obtained using $\boldsymbol{v}'=\boldsymbol{X}^{-1} \boldsymbol{v}$ that are mutually orthogonal (see Appendix~\ref{sec:a2} for details). Note that since the transition matrix, $\boldsymbol{W}$ is in general non-Hermitian due to the broken detailed balance,  the expansion in Eq.~(\ref{time evolution}) is possible only in the case when its eigenvectors $\boldsymbol{v}$'s are mutually independent of each other. The criterion to check that its eigenvectors are independent is that the determinant of the matrix $\boldsymbol{X}$ (see Appendix~\ref{sec:a2}) formed by taking its columns as the eigenvectors $\boldsymbol{v}$ of $\boldsymbol{W}$ is non-zero. If there is a case where the determinant is zero, then the eigenvectors $\boldsymbol{v}$ do not form a basis. In that scenario, matrix $\boldsymbol{W}$ cannot be diagonalized and hence orthogonal conjugate eigenvectors $\boldsymbol{v}' $ also do not exist.

If the transition rates satisfy detailed balance, it is guaranteed that all the eigenvalues are real \cite{schnakenberg1976network}. On the other hand, for non-equilibrium dynamics with broken detailed balance, the eigenspecturm can be mixed with the presence of both real and complex eigenvalues (see Appendix~\ref{sec:simple 3 state without detailed balance}). Here too, we observe that the eigenvalues turn out to be complex as one varies the activity parameter $\gamma$. Figure~\ref{eigenspectrum} illustrates such a transition in the eigenspectrum as a result of the change in the active parameters $\epsilon$ and $\gamma$.  Figure~\ref{eigenspectrum}(b) further describes the pattern in the eigenspectrum as $\epsilon$ and $\gamma$ are varied. The eigenspectrum is purely real for the combination of very small $\epsilon$ while fixing $\gamma$ either small or large. On the contrary, the minimum number (two) of real eigenvalues occurs  at larger $\epsilon$, for a given $\gamma$.

\begin{figure}
\centering
\includegraphics[width=\columnwidth]{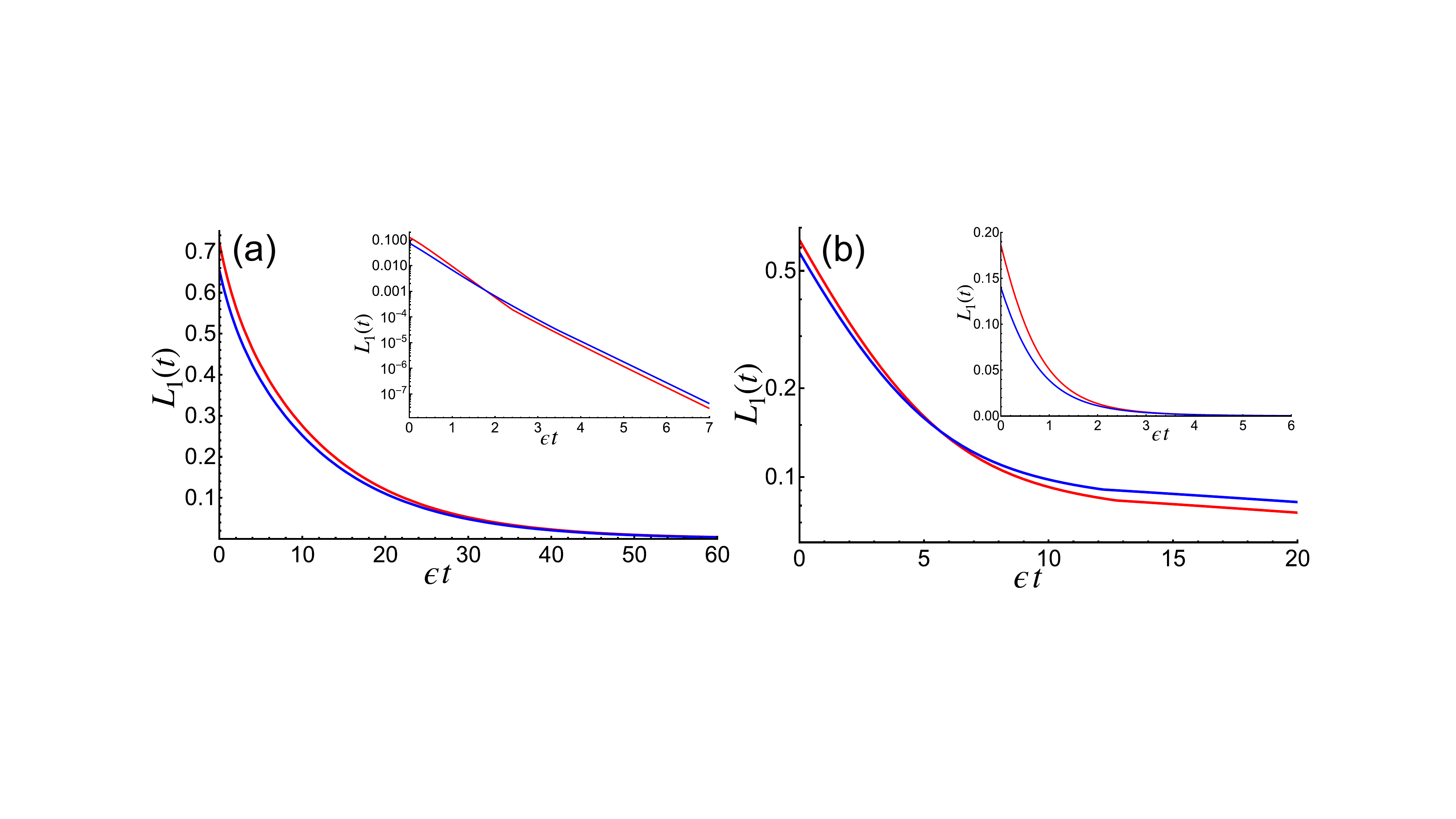} 
\caption{\label{activity induced mpemba}\textbf{Activity induced and suppressed Mpemba effect}. Panel (a): The Mpemba effect is absent in the passive case ($\epsilon/R_{12}(T_b)=\gamma/R_{12}(T_b)=0$) while the presence of activity ($\epsilon/R_{12}(T_b)=40.17$ and $\gamma/R_{12}(T_b)=80.34$) induces the Mpemba effect (shown in inset). The model parameters are: $E_1=0$, $E_2= T_b$, $E_3=3 T_b$, $B_{12}=B_{21}=4 T_b$, $B_{13}=B_{31}=5 T_b$, $B_{23}=B_{32}=4 T_b$, $T_b=0.1$, $T_h/T_b=40$ and $T_c/T_b=10$. Panel (b) shows activity suppressed Mpemba effect (shown in inset) while it is present in the passive case. Here, the active rates are $\epsilon/R_{12}(T_b)=403.43$ and $\gamma/R_{12}(T_b)=806.86$. The model parameters in this case are: $E_1=2 T_b$, $E_2=3 T_b$, $E_3=5 T_b$, $B_{12}=B_{21}=9 T_b$, $B_{13}=B_{31}=6 T_b$, $B_{23}=B_{32}=9 T_b$, $T_b=0.1$, $T_h/T_b=30$ and $T_c/T_b=10$. 
}
\end{figure}

\section{The Mpemba effect}\label{sec:mpemba effect}

To interpret the Mpemba effect, let us consider two systems: first one $A$, initially in a steady state corresponding to a \textit{hot} temperature $T_h$ and second one $B$, initially in a different steady state corresponding to \textit{cold} temperature $T_c$ where  $T_h > T_c$.  These initial steady states are denoted by $\boldsymbol{\pi}(T_h)$ and $\boldsymbol{\pi}(T_c)$ respectively. Now imagine that both $A$ and $B$ are quenched at once to a common bath temperature, $T_b$, where $T_h > T_c>T_b$. Eventually, both of them will relax to the common distribution  $\boldsymbol{\pi}(T_b)$ given long enough time. The Mpemba effect is said to exist if $A$ reaches the final steady state faster than $B$ during the transient/relaxation process. This relaxation can be quantified in terms of the \textit{distance function} $D[\boldsymbol{\mathcal{P}}(t),\boldsymbol{\Pi}(T_b)]$ which measures the instantaneous distance of the state-wise occupation probabilities $\boldsymbol{\mathcal{P}}(t)=(P_1, P_2, P_3)^{\mathbb{T}}$ from their respective final steady state distributions $\boldsymbol{\Pi}(T_b)=(\pi_1, \pi_2, \pi_3)^{\mathbb{T}}$. There are many well-adapted measures that exist in the literature namely the norm $L_1(t)=\sum_{i=1}^3 |P_i(t)-\pi_i(T_b)|$ and Kullback-Leibler (KL) divergence  $KL(t)=\sum_{i=1}^3 P_i (t) \ln(P_i(t)/\pi_i)$.  Thus, if one has $D[\boldsymbol{\Pi}(T_h), \boldsymbol{\Pi}(T_b)]>D[\boldsymbol{\Pi}(T_c), \boldsymbol{\Pi}(T_b)]$ initially for $T_h > T_c$ followed by $D[\boldsymbol{\Pi}(T_h), \boldsymbol{\Pi}(T_b)]<D[\boldsymbol{\Pi}(T_c), \boldsymbol{\Pi}(T_b)]$ at a later time, we infer that the Mpemba effect exists. 

\section{Activity harnessed Mpemba effect}\label{sec:activity harnessed mpemba effect}

\begin{figure}
\centering
\includegraphics[width=\columnwidth]{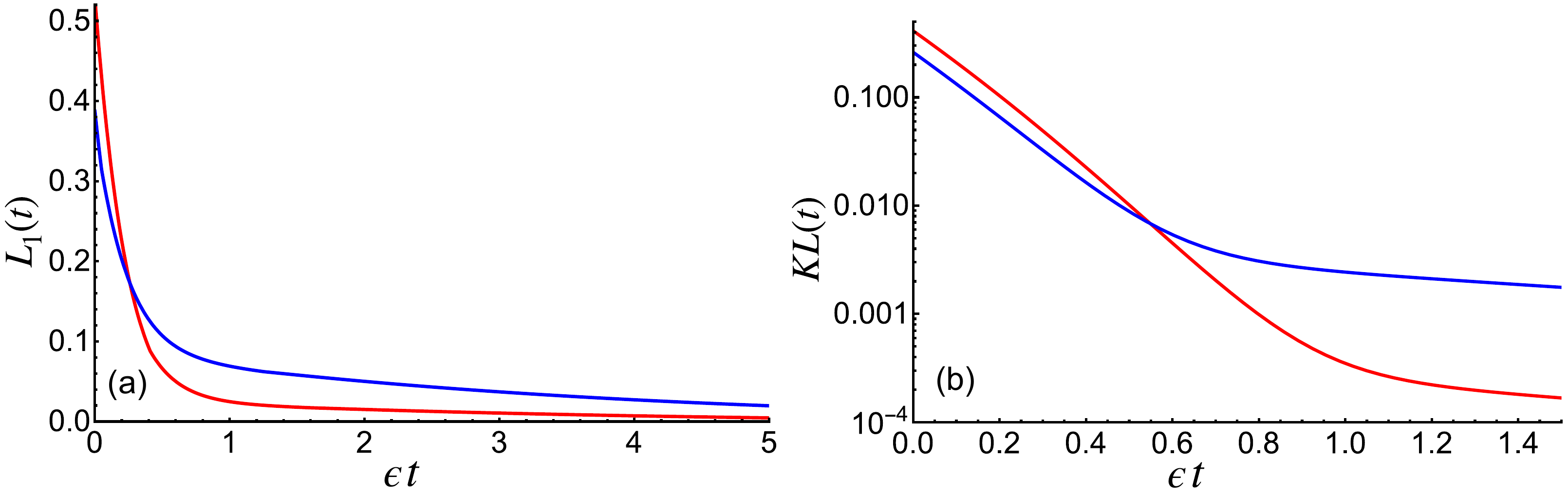} 
\caption{\label{combined normal mpemba}\textbf{Inference of the Mpemba effect is independent of the choice of the distance measure.} Illustration of the Mpemba effect using: (a) $L_1$-norm, and (b) Kullback-Leibler (KL) divergence. Here, the model parameters: $E_1=0$, $E_2= T_b$, $E_3=5 T_b$, $B_{12}=B_{21}=6 T_b$, $B_{13}=B_{31}=5 T_b$, $B_{23}=B_{32}=4 T_b$, $\gamma/R_{12}(T_b)=593.65$, $\epsilon/R_{12}(T_b)=148.41$, $T_b=0.1$ whereas the initially $hot$ and $cold$  temperatures are chosen to be: $T_h/T_b=30$ and $T_c/T_b=8$ respectively.}
\end{figure}

To highlight the effect of activity, we first set 
$\epsilon=\gamma=0$ and consider the passive discrete three-state Markov chain \cite{Lu-raz:2017}. The identification of the Mpemba effect in this case can often be made from the behaviour of the coefficient $|a_2(T,T_b)|$ [as in Eq.~(\ref{time evolution})], namely by checking whether $|a_2(T_c,T_b)|>|a_2(T_h,T_b)|$ or by measuring the distance functions \cite{Lu-raz:2017}. In this set-up, let us now choose a set of parameters $R_{ij}$-s such that the Mpemba effect is absent. However, as the activity is introduced by setting $\epsilon,\gamma >0$, we notice a spontaneous emergence of the Mpemba effect [see Fig.~\ref{activity induced mpemba}]. We also observe a converse effect where the introduction of activity suppresses an already existing Mpemba effect (Fig.~\ref{activity induced mpemba}). Quite distinctly, this behavior remains robust under other distance measures. To see this, we measure the relaxation in terms of two popular distance measures: $L_1$-norm and KL divergence. Figure~\ref{combined normal mpemba} shows, for a particular choice of the model parameters, that both the measures infer the existence of the Mpemba effect. For convenience, we use $L_1$-norm in all the instances unless otherwise stated. It is worthwhile to note that the measure $|a_2(T,T_b)|$ is in general not an appropriate measure to describe the Mpemba effect in systems with broken detailed balance and complex eigenspectrum. This is because such systems do not necessarily show monotonic relaxations and consequently, \textit{multiple crossings} between the relaxation trajectories prepared from different initial conditions can emerge, as will be discussed later (also see \cite{biswas2023measure,chatterjee2024multiple,PhysRevE.103.032901}). This is in contrast with monotonic relaxations (as in the case of $\epsilon=\gamma=0$) where the contribution of the second eigenmode reflected by $|a_2(T,T_b)|$ becomes crucial (see the discussion in Appendix~\ref{sec:real}).

%
%
%
%

The qualitative reasoning of the occurrence of the Mpemba effect in the energy landscape with multiple local minima is given in terms of the effective trapping of the initially prepared colder system in the metastable states leading to its slower relaxation. In the context of the current model, a passive random walker initiated at a hot temperature does not experience any trap and explores the energy landscape much faster than the initially prepared colder system. On the contrary, an active random walker, be it initiated at a hot or cold temperature or even for a significant barrier height of the landscape, may not experience any metastable state for intermediate values of its activity or persistence as it can now explore the phase space with additional frequency characterized by the active rates. Such realizations impede the Mpemba effect in the presence of activity. However, the activity can also act otherwise. For example, the active random walker can experience momentary trap in the energy landscape due its persistence in a particular state leading to an activity induced Mpemba relaxation. Summarizing, the activity induced behavior is a complex convolution of the initial temperatures; the active and diffusive rates of the particle; barrier heights of the underlying energy landscape and the activated escape rate from the metastable states \cite{woillez2019activated,woillez2020nonlocal}.

\begin{figure}
\centering
\includegraphics[width=\columnwidth]{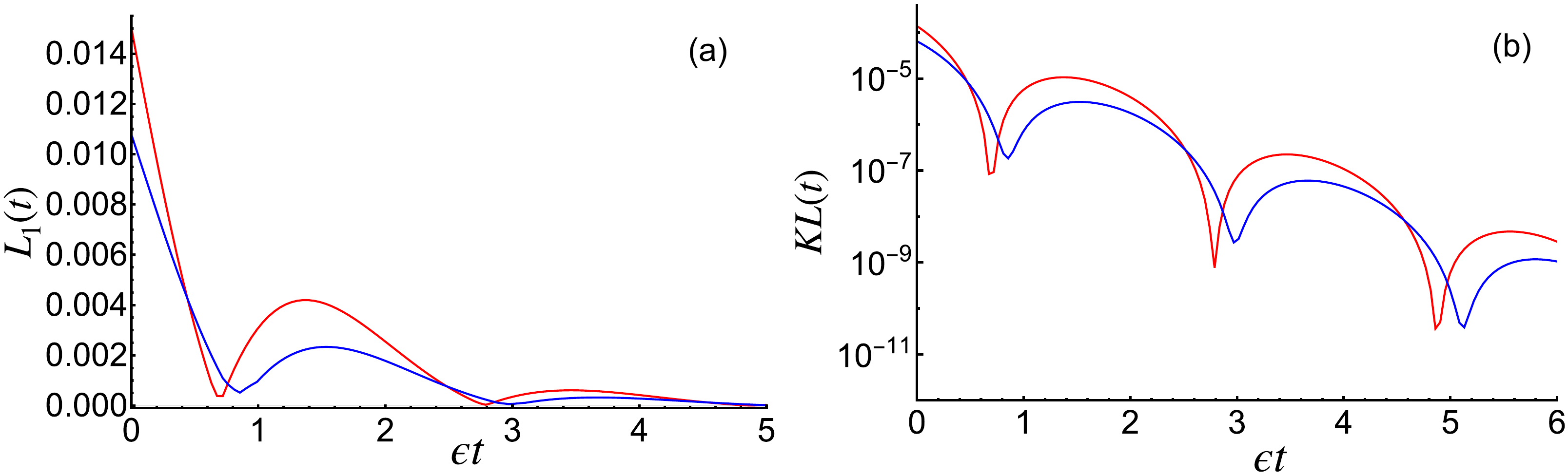} 
\caption{\label{oscillatory mpemba diff measures} Illustration of the \textbf{oscillatory Mpemba effect} across different distance measures: (a) the $L_1$-norm, and (b) the Kullback-Leibler (KL) divergence for a system with broken detailed balance. The complex eigenspectrum can lead to oscillatory relaxations and distance measures. Here, the model parameters are: $E_1=0$, $E_2=T_b$, $E_3=3 T_b$, $B_{12}=B_{21}=4 T_b$, $B_{13}=B_{31}=5 T_b$, $B_{23}=B_{32}=4 T_b$, $\gamma/R_{12}(T_b)=160.68$, $\epsilon/R_{12}(T_b)=180.77$, $T_b=0.1$ whereas the temperatures of the hot and cold systems are chosen to be: $T_h/T_b=50$ and $T_c/T_b=5$ respectively.}
\end{figure}

\begin{figure*}
\centering
\includegraphics[width=\textwidth]{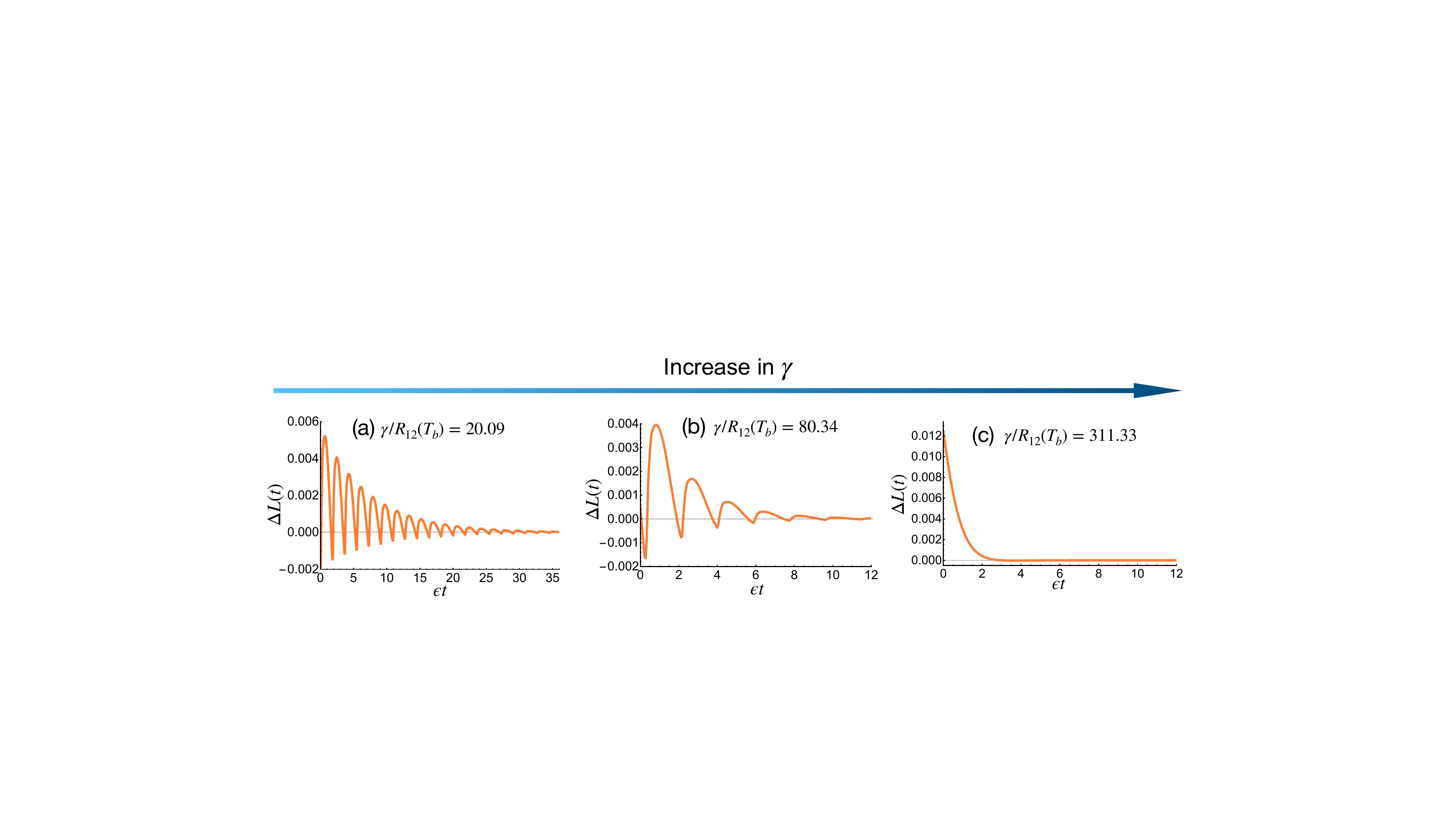} 
\caption{\label{diff of L1}\textbf{Oscillatory Mpemba effect}
characterized by the \textit{zero crossings} in $\Delta L_1(t):= L^{hot}_1(t)-L^{cold}_1(t)$ for two initially prepared hot and cold systems as they relax to a common steady state. We measure $\Delta L_1(t)$ as a function of time for varying flipping rate $\gamma/R_{12}(T_b)$ which shows a pronounced effect of the activity. The oscillations tend to go down with increased $\gamma/R_{12}(T_b)$ as the active walker becomes a diffusive one in this limit \cite{jose2022active}. Here, we have set  $\epsilon/R_{12}(T_b)=180.77$ throughout the analysis.} 
\end{figure*}

\section{Oscillatory Mpemba effect}\label{sec:oscillatory mpemba}
It is evident from Fig.~\ref{eigenspectrum} that the eigenvalues of the transition matrix  $\boldsymbol{W}(T_b)$  can be \textit{complex}. As a result, both the time evolution of the probability distribution and the distance measures will have oscillatory terms. Figure~\ref{oscillatory mpemba diff measures} illustrates the time evolution of two different distance measures, namely $L_1$ and KL, for two systems $A$ and $B$ that are initiated at two different temperatures $T_h$ and $T_c$ respectively with $T_h>T_c$. We show that unlike the previous results of the Mpemba effect where the distance measures decay monotonically with time, here the distance measures show non-monotonic oscillatory-like decay across all the measures. As a result, the trajectories of the initially \textit{hot} and \textit{cold} system cross multiple times (hence referred as the \textit{Oscillatory Mpemba effect}) and $|a_2(T,T_b)|$-criterion is not suitable to depict such situation. Notably, the distance measures such as $L_1(t)$ and $KL(t)$ can contain superposition of multiple frequencies and not a unique one, which we discuss in detail in Sec.~\ref{sec:comparison with damped oscillator}. 

To have a better representation for the oscillations, we plot the difference of the $L_1$-norm for the hot and the cold system namely $\Delta L_1(t):=L^{hot}_1(t)-L^{cold}_1(t)$ as  a function of time $t$ for different activity rates $\gamma$ keeping $\epsilon$ fixed as shown in Fig.~\ref{diff of L1}. The plots show that $\Delta L_1$ crosses the origin multiple times before it asymptotically relaxes to zero. The number of oscillations decreases as function of $\gamma$ for a fixed $\epsilon$ and thus, for the higher values of $\gamma$, we can observe a re-entrant diffusive like behavior in the relaxation of the trajectories. This is not surprising as in the large flipping rate $\gamma$ limit, the active walker behaves like a passive one as can also be seen from the effective diffusion constant \cite{jose2022active,malakar2018steady}.

\begin{figure}
\centering
\includegraphics[width=\columnwidth]{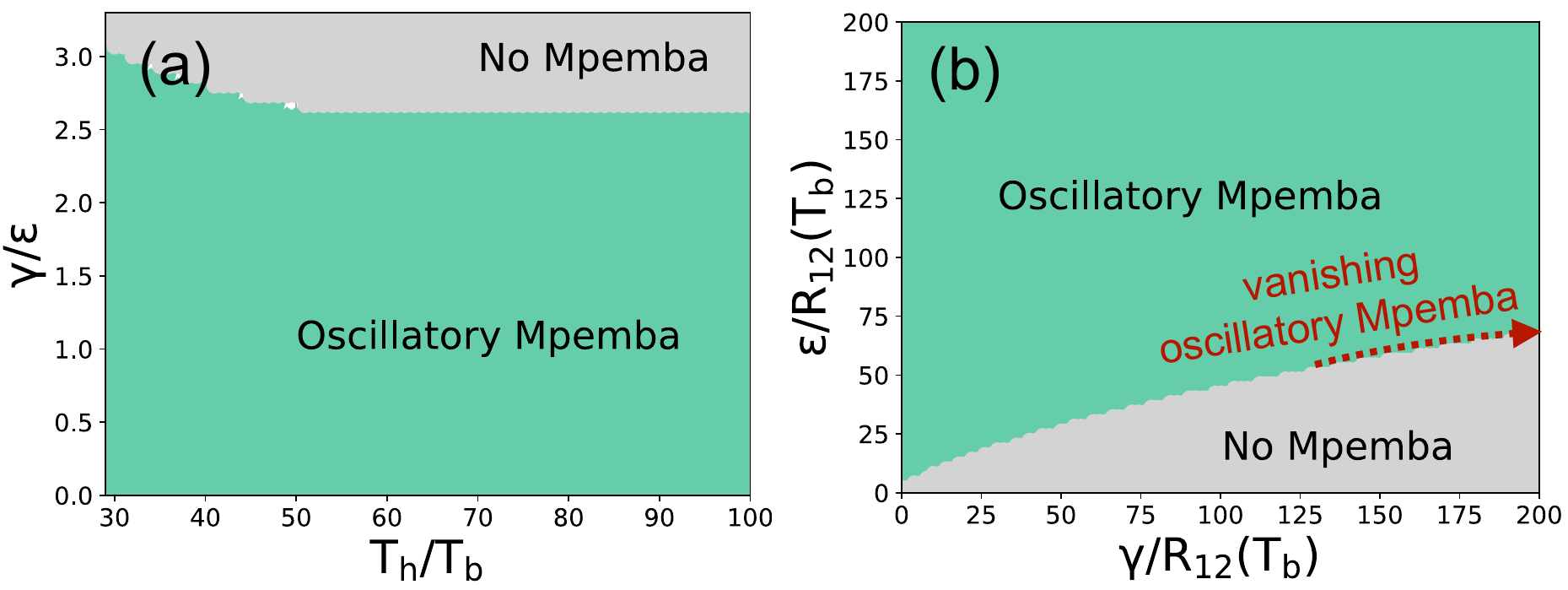} 
\caption{\label{phase_diagram} Phase diagram for the \textbf{oscillatory Mpemba effect}. Panel (a) illustrates the phase space region that shows the presence or absence of the Mpemba effect as the ratio of the activity $\gamma/\epsilon$ is varied along with $T_h$ keeping the other parameters fixed: $E_1=0$, $E_2=T_b$, $E_3=3 T_b$, $B_{12}=B_{21}=4 T_b$, $B_{13}=B_{31}=5 T_b$, $B_{23}=B_{32}=4 T_b$ and $T_b=0.1$. Panel (b) illustrates the variation in the phase space in the $\epsilon-\gamma$ plane setting $T_h/T_b=40$ and other parameters as in panel (a). The region for \textit{non-oscillatory Mpemba effect} is seen to expand with large $\gamma$. Here, we have set $T_c/T_b=10$. 
}
\end{figure}

\begin{figure*}
\centering
\includegraphics[width=\textwidth]{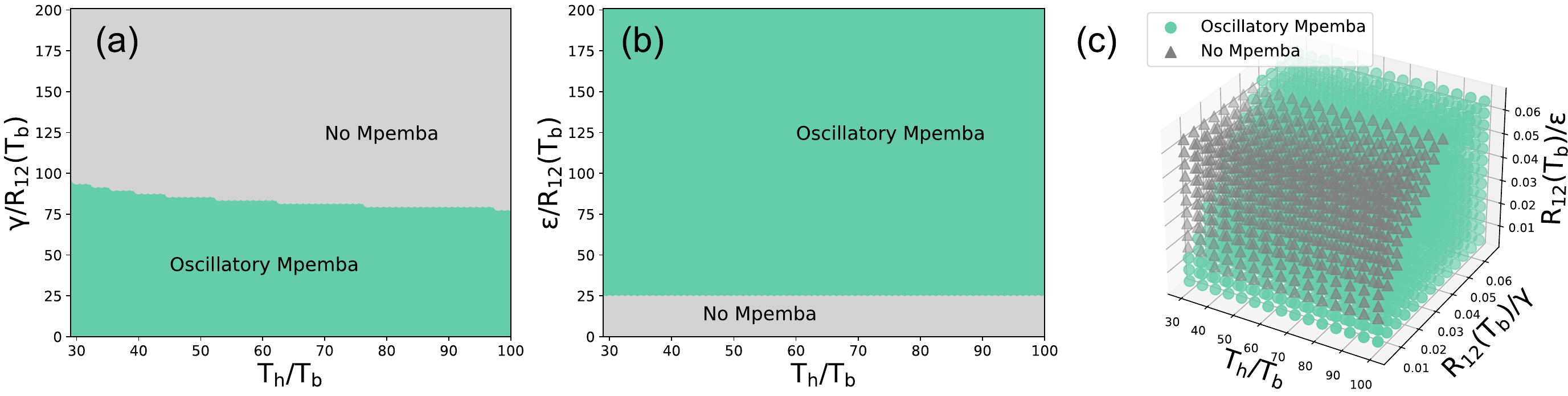} 
\caption{\label{combined phase diagram gamma and epsilon vs Th}\textbf{Phase diagram of the active Mpemba effect}. Panel (a) illustrates the phase space region that shows the presence or absence of the Mpemba effect as the flipping rate $\gamma$ is varied along with $T_h$, the temperature of the hot system, whereas the other parameters are kept fixed at $E_1=0$, $E_2=T_b$, $E_3=3 T_b$, $B_{12}=B_{21}=4 T_b$, $B_{13}=B_{31}=5 T_b$, $B_{23}=B_{32}=4 T_b$, $\epsilon/R_{12}(T_b)=40.17$ and $T_b=0.1$. Panel (b) illustrates the variation in the phase space in the $\epsilon$ versus $T_h$ plane but now keeping $\gamma/R_{12}(T_b)=40.17$ fixed and the other parameters are same as in panel (a).  Panel (c) illustrates the phase space of the oscillatory Mpemba effect in terms of relevant timescales of activity parameters: $1/\gamma$ and $1/\epsilon$  as a function of the variation in $T_h$. The other model parameters are kept fixed as per (a) and (b), and the temperature of the initially cold system in all the cases, are chosen to be $T_c/T_b=10$.}
\end{figure*}

We delve deeper by analyzing the variation in active parameter space and the consequence to the oscillatory Mpemba effect. To this end, first we focus on two phase diagrams spanned by -- (a) ratio of the active rates $\gamma/\epsilon$ and the temperature of the hot reservoir $T_h$, and (b) active rates $\epsilon$ and $\gamma$, keeping the other parameters fixed respectively (see Fig.~\ref{phase_diagram}). In the former case, the phase boundary remains insensitive to $T_h$ following an initial slow variation thus maintaining two distinct regions. In the latter case, the phase boundary changes if we vary $\epsilon$ and $\gamma$. It becomes evident that as $\gamma$ is increased, the domain of oscillations will gradually shrink indicating a vanishing Mpemba effect, as was indicated in Fig.~\ref{diff of L1}. In addition, now we discuss few additional plots [see Fig.~\ref{combined phase diagram gamma and epsilon vs Th}] that describes the behavior of the activity on the anomalous cooling but in the following planes of phase space: (a) variation of $\gamma$ with $T_h$, (b)  variation of $\epsilon$ with $T_h$, and (c) simultaneous variation of $\gamma$, $\epsilon$ and $T_h$, while keeping every other parameter of the model that is not taken as a variation in the plane of the diagram is kept fixed. Figure~\ref{combined phase diagram gamma and epsilon vs Th}(a) shows that for a fixed value of the model parameters, there exists a critical $\gamma$ that separates the region of oscillatory Mpemba with no Mpemba region, and with the increase in $\gamma$, the area of the oscillatory Mpemba region decreases as discussed earlier in Fig.~\ref{phase_diagram}. However, the behavior of the phase space variation is different in $\epsilon-T_h$ plane as shown in Fig.~\ref{combined phase diagram gamma and epsilon vs Th}(b). Here, the critical value of $\epsilon$ that separates the region of oscillatory Mpemba with no Mpemba, remains the same for the chosen values of other parameters of the model. The combined analysis with the simultaneous variation of $\gamma$ and $\epsilon$ with $T_h$ is described in Fig.~\ref{combined phase diagram gamma and epsilon vs Th}(c). However, the phase space now depicts the relevant timescales of the activity, i.e., $1/\gamma$ and $1/\epsilon$, that results in the oscillatory Mpemba effect  as a function of the variation in $T_h$.
These comprehensive phase diagrams provide a cue to asses the relevant active timescales in probing the oscillatory effect.

\section{\label{sec:comparison with damped oscillator} Presence of multiple frequency of oscillations in the oscillatory Mpemba effect}
The oscillatory Mpemba effect or equivalently the oscillatory behavior in $L_1(t)$ (or $\Delta L(t)$) [see Fig.~\ref{oscillatory mpemba diff measures} and Fig.~\ref{diff of L1}] can be presumed to have certain resemblance with that of a simple damped oscillator. In this section, we make an attempt to understand whether there is a similarity in the oscillatory behavior in $L_1(t)$ with that of a simple damped oscillator. For that let us expand the expression for the measure $L_1(t)$ in terms of eigenmodes: $L_1(t)=\sum_{i=1}^3 |P_i(t)-\pi_i(T_b)|$,
where $P_i(t)$ denotes the occupation probability of site $i$ at time $t$ and $\pi_i(T_b)$ denotes its final steady state probability at temperature $T_b$. The probabilities $P_i(t)$ for sites $i=1, 2, 3$ are obtained from the total probability vector $\boldsymbol{P}=(P_1, P_2, P_3, P^*_1, P^*_2, P^*_3)^{\mathbb{T}}$ [see Eq. (\ref{time evolution})] by taking the first three components and their expansions in terms of eigenmodes can be written following the expansion of $\boldsymbol{P}$ [see Eq.~(\ref{time evolution})] as follows:
\begin{flalign}
P_j(t)=\pi_j(T_b)+\sum^{6}_{i=2}a_i (T, T_b) v_{ij} e^{\lambda_i t},\quad j=1,2,3, \label{eqn for occup prob}
\end{flalign}
where $v_{ij}$ denotes the $j^{th}$ component of $i^{th}$ eigenvector $\boldsymbol{v}_i$ of the transition matrix $\boldsymbol{W}(T_b)$ and $\lambda_i$ are its eigenvalues. For simplicity, while writing $P(t)$ from $\boldsymbol{P}(T,t)$ using Eq.~(\ref{time evolution}), we have omitted $T$ from its parenthesis.  Putting the expansion of Eq.~(\ref{eqn for occup prob}) in the expression for $L_1(t)$, we obtain

\begin{equation}
\begin{split}
L_1(t)&=\vert\sum^{6}_{i=2}a_i (T, T_b) v_{i1} e^{\lambda_i t} + \sum^{6}_{i=2}a_i (T, T_b) v_{i2} e^{\lambda_i t} \\
&\ + \sum^{6}_{i=2}a_i (T, T_b) v_{i3} e^{\lambda_i t}|=|\sum^{6}_{i=2}a_i (T, T_b) e^{\lambda_i t} \sum^{3}_{j=1} v_{ij} \vert.
\end{split}
\end{equation}

We recall that in the eigenspectrum of the transition matrix $\boldsymbol{W}(T_b)$, the eigenvalue $\lambda_1=0$ corresponds to the steady state $\boldsymbol{\pi}(T_b)$ and the remaining eigenvalues $\lambda_i$, $i=2, \ldots 6$ are in general complex with negative real parts. Since complex eigenvalues occur in pair, of the remaining eigenvalues there \textit{can be} at most 4 complex eigenvalues, i.e., two pairs of complex numbers and in that case, the remaining one eigenvalue has to be real. 

Consider this generic case where the eigenvalue $\lambda_2$ is real while $\lambda_{3,4}=\beta_r \pm \mathrm{i} \beta_c$ and $\lambda_{5,6}=\delta_r \pm \mathrm{i} \delta_c$ are complex eigenvalues. With this consideration and dropping the arguments of coefficients $a_i$ for simplicity, we can write $L_1(t)$ as

\begin{widetext}
\begin{equation}
\begin{split}
L_1(t)&=|  a_2  e^{\lambda_2 t} \sum^{3}_{j=1} v_{2j} + a_3 e^{(\beta_r + \mathrm{i} \beta_c)t } \sum^{3}_{j=1} v_{3j} + a_4 e^{(\beta_r - \mathrm{i} \beta_c)t } \sum^{3}_{j=1} v_{4j}   + a_5 e^{(\delta_r + \mathrm{i} \delta_c)t } \sum^{3}_{j=1} v_{5j} + a_6 e^{(\delta_r - \mathrm{i} \delta_c)t } \sum^{3}_{j=1} v_{6j} |\\
&=| a_2 \sum^{3}_{j=1} v_{2j}  e^{\lambda_2 t} + e^{\beta_r t} \big[ (a_3 \sum^{3}_{j=1} v_{3j}+a_4 \sum^{3}_{j=1} v_{4j}) \text{cos}(\beta_c t) + \mathrm{i} (a_3 \sum^{3}_{j=1} v_{2j}-a_4 \sum^{3}_{j=1} v_{2j}) \text{sin}(\beta_c t) \big] \\
&\quad   + e^{\delta_r t} [ (a_5 \sum^{3}_{j=1} v_{5j}+a_6 \sum^{3}_{j=1} v_{6j}) \text{cos}(\delta_c t) + \mathrm{i} (a_5 \sum^{3}_{j=1} v_{5j}-a_6 \sum^{3}_{j=1} v_{6j}) \text{sin}(\delta_c t) ] |\\
&=|\big( a_2 \sum^{3}_{j=1} v_{2j}  e^{\lambda_2 t} + A_{\beta} e^{\beta_r t} \text{cos}(\beta_c t +\phi_{\beta}) + A_{\delta} e^{\delta_r t} \text{cos}(\delta_c t +\phi_{\delta}) \big)|,
\end{split}
\end{equation}
\end{widetext}
where $A_\beta$, $A_\delta$ and $\phi_\beta$, $\phi_\delta$ denote the respective effective amplitudes and phases for the two modes of damped oscillations. Thus, in this case the \textit{oscillation in the quantity $L_1(t)$ is in effect a superposition of ``two frequencies}" corresponding to the imaginary part of the two pairs of complex eigenvalues.

On the other hand, a simple damped oscillator has a ``\textit{single frequency}" (say, $\nu_0$). Thus, in general there exists no \textit{one to one mapping} between the oscillations in the quantity $L_1(t)$ [likewise the quantity $\Delta L(t)=L^{hot}_1(t)-L^{cold}_1(t)$] and a simple damped oscillator. Evidently, in certain cases, when there exists only one pair of complex eigenvalues, such a restrictive correspondence can be made. As an illustration, we recall the case considered in Fig.~\ref{oscillatory mpemba diff measures} -- it should be noted that the eigenspectrum in this case has two pairs of complex eigenvalues and the same can be confirmed also from eigenvalue phase diagram Fig.~\ref{eigenspectrum}(b). Evidently, there is no single frequency in this case as the preceding discussion suggests.

However, a particular case can be considered where $\beta_r$ happens to be the largest eigenvalue, in terms of its magnitude of its real part of the eigenvalue. In the large time limit, only the non-zero largest eigenvalue shall dominate and  we can map the oscillatory decaying coefficient of the measure  $\Delta L(t)$ [or $L_1(t)$] with the $a_2$ coefficient [see Eq.~(\ref{approx time evolution})] (which is also related with the largest non-zero eigenvalue) as
\begin{align}
a_2(T,T_b)=\frac{A_{\beta}  \text{cos}(\beta_c t +\phi_{\beta})}{\sum^{3}_{j=1} v_{2j}}.
\end{align}


\section{\label{sec:multiple crossing with real eigenspectrum}Multiple Mpemba effect even with real eigenspectrum}

\begin{figure}
\centering
\includegraphics[width=\columnwidth]{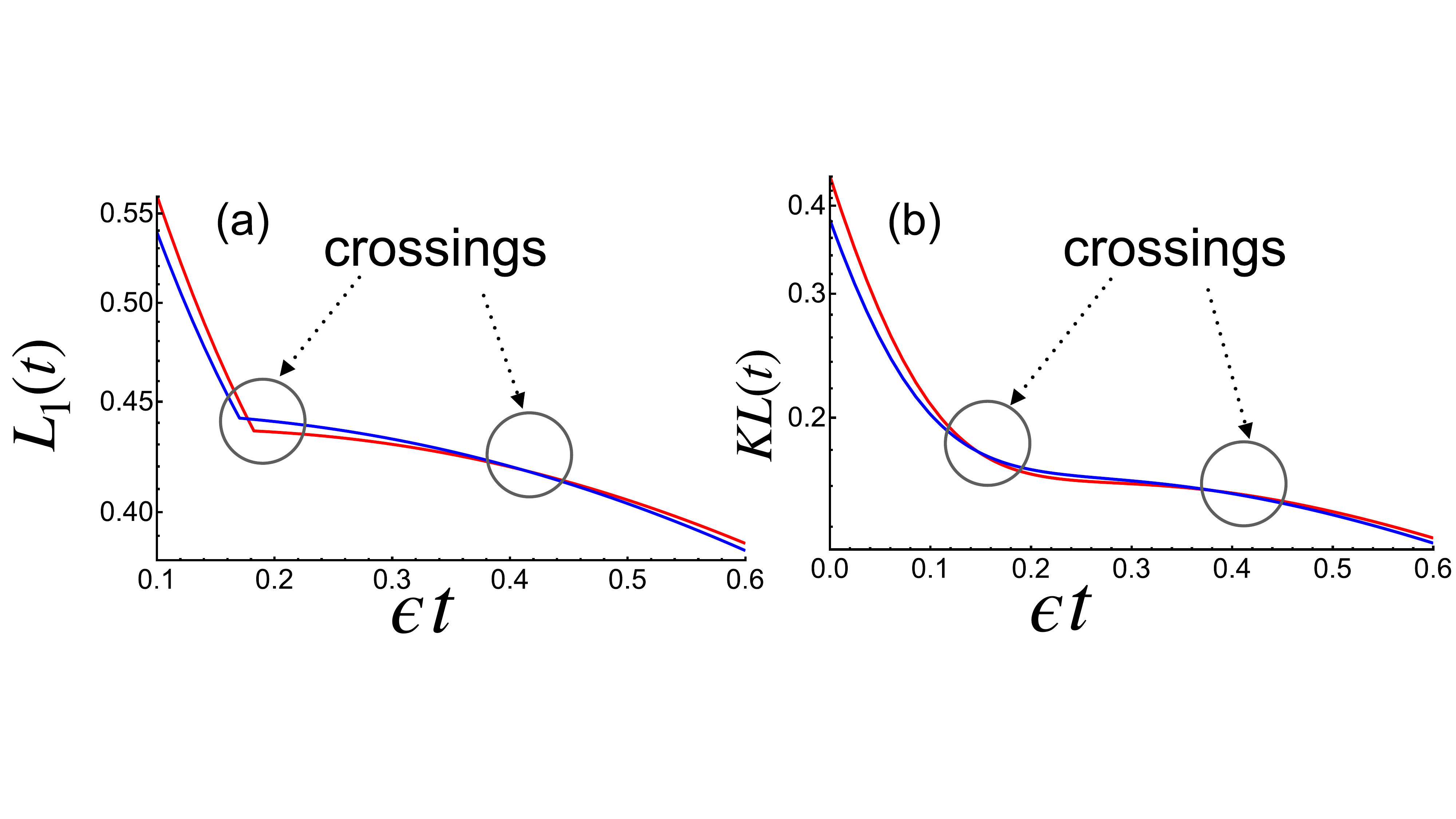} 
\caption{\label{non-degenerate case multiple crossing}Multiple crossings in the relaxation trajectories can emerge even with \textit{real eigenspectrum} of the active Markov chain (with $\gamma/R_{12}(T_b)=\epsilon/R_{12}(T_b)=148.41$). The result is independent of distance measures used as shown here with: (a) $L_1$-norm, and (b) Kullback-Leibler (KL) divergence. The results shown are for model parameters: $E_1=0$, $E_2=T_b$, $E_3=3 T_b$, $B_{12}=B_{21}=6 T_b$, $B_{13}=B_{31}=T_b$, $B_{23}=B_{32}=6 T_b$, $T_b=0.1$, whereas the temperatures of the hot and cold systems are chosen to be: $T_h/T_b=100$ and $T_c/T_b=10$ respectively.}
\end{figure}

 \textit{Is it always ensured to have a single crossing of trajectories with real eigenspectrum?} We consider one such case but find that the distance functions can cross \textit{twice} before they asymptotically merge with each other as illustrated in Fig.~\ref{non-degenerate case multiple crossing}. Since there are even crossings, the hotter system lags behind the initially colder system leading to no net Mpemba effect in the large time. The same conclusion is drawn from the measure $|a_2(T,T_b)|$ as well [see Fig.~\ref{supp: a2 real eigensystem multiple crossing}]. Note that such a scenario is also observed in the non-equilibrium model of driven granular gases~\cite{PhysRevE.103.032901,biswas2023measure} and molecular gases in the presence of non-linear drag~\cite{megias2022thermal} (albeit not universally but depending on the choice of the distance measure). Generically, this behavior can be attributed to the preparation of the systems in non-equilibrium conditions as is the case for the active systems here.

\section{\label{Conclusion}Conclusion}
In this paper, we have investigated the role of broken detailed balance in the anomalous relaxation phenomena. We unveil that the broken detailed balance conferred by the activity in a Markov chain system can showcase a myriad of unique relaxation phenomena such as the activity -induced and -suppressed Mpemba effect, and oscillatory Mpemba effect where relaxation trajectories cross multiple times. The latter is owing to the emergence of the complex eigenspectrum of the transition matrix due to the breakdown of detailed balance in the presence of activity. 

Qualitatively, some of these phenomena can be reasoned with a heuristic description of `activity induced exploration on the energy landscape'. For instance, a \textit{cold} walker can overcome metastabilities, kinetic traps or bottlenecks with the aid of active rates suppressing the Mpemba effect. However, activity can also act in favour of the Mpemba effect by trapping the walker of the initially colder system in a particular state with strong persistence. This would lead to the activity induced Mpemba effect. An intermediate persistence renders the walker to switch its state intermittently and thus, one would observe a non-monotonic relaxation behavior.

While it is known that the active fluctuations can lead to an extensive variety of chemical and biological functions in living systems, its direct role in the quenching and relaxation dynamics pertaining to the Mpemba effect is not well understood. Drawing inspiration from this rigorous analysis of an active Markov chain, we posit that some of these ideas will be important to explore in the coarse-grained living systems such as fluctuating enzymes \cite{qian2007phosphorylation,moffitt2014extracting}, proteins, or molecular machines \cite{kolomeisky2015motor} that continuously dissipate energy with the surrounding or exchange energy with the other molecules \cite{gnesotto2018broken}.

It would be interesting to see whether experimental setups can be designed to capture 
such Mpemba like non-equilibrium phenomena from the transport and relaxation of bio-molecules through the engineered structures and cell biological systems \cite{chou2011non} or even in simple active colloidal systems \cite{zottl2016emergent}. We believe that a synergy between the theory and experiment can provide a deeper conceptual understanding of this non-equilibrium phenomena.


\section{Acknowledgement}
 AP thanks Amit Kumar Chatterjee and Hisao Hayakawa for useful discussions and for sharing relevant references on  the quantum Mpemba effect. AP also thanks Udo Seifert for insightful discussions. AP acknowledges research support from the DST SERB Start-up Research Grant Number
SRG/2022/000080 and the Department of Atomic Energy,
Government of India. AP also acknowledges the research support from the
International Research Project (IRP) titled “Classical
and quantum dynamics in out of equilibrium systems”
by CNRS, France.

\appendix

\section{\label{sec:solution}Time evolution of occupation probabilities on a 3-state active Markov chain}
This section discusses the time evolution of  the occupation probabilities of a random walker on a three state Markov chain. The walker can hop  from site $j$ to site $i$  and vice-versa with an intrinsic rate $R_{ij}$ that satisfy detailed balance. Furthermore, the walker is active and we characterize the activity of the walker in terms of the flipping rate $\gamma$ between its internal states $\uparrow$ and $\downarrow$. Depending on the current internal state ($\uparrow$ and $\downarrow$), the walker hops either to the clockwise or the anti-clockwise site at a rate $\epsilon$. Thus, $\gamma$ is the effective persistence rate and  $\epsilon$ can be viewed as an effective velocity while in a given state \cite{jose2022active}.

Let $p_{i,\uparrow}$ denotes the joint occupation probability of the walker to be in site $i$ and $\uparrow$ state and vice-versa for $p_{i,\downarrow}$, where $i=1,2,3$ denotes the sites. With these definitions, the time evolution of the joint occupation probabilities are given by:
\begin{widetext}

\noindent
\textit{For site 1:}
\begin{align}
\begin{split}
\frac{d p_{1,\uparrow}}{d t}=(R_{12}+\epsilon) p_{2,\uparrow} + (R_{13}-\epsilon) p_{3,\uparrow}
-(R_{21}+R_{31}) p_{1,\uparrow} -\gamma p_{1,\uparrow}+ \gamma p_{1,\downarrow},\\
\frac{d p_{1,\downarrow}}{d t}=(R_{12}-\epsilon) p_{2,\downarrow} + (R_{13}+\epsilon) p_{3,\downarrow} 
-(R_{21}+R_{31}) p_{1,\downarrow} -\gamma p_{1,\downarrow}+ \gamma p_{1,\uparrow}.
\end{split}
\end{align}
\noindent
\textit{For site 2:}
\begin{align}
\begin{split}
\frac{d p_{2,\uparrow}}{d t}=(R_{21}-\epsilon) p_{1,\uparrow} + (R_{23}+\epsilon) p_{3,\uparrow}
-(R_{12}+R_{32}) p_{2,\uparrow} -\gamma p_{2,\uparrow}+ \gamma p_{2,\downarrow},\\
\frac{d p_{2,\downarrow}}{d t}=(R_{21}+\epsilon) p_{1,\downarrow} + (R_{23}-\epsilon) p_{3,\downarrow} 
-(R_{12}+R_{32}) p_{2,\downarrow} -\gamma p_{2,\downarrow}+ \gamma p_{2,\uparrow}.
\end{split}
\end{align}

\noindent
\textit{For site 3:}
\begin{align}
\begin{split}
\frac{d p_{3,\uparrow}}{d t}=(R_{31}+\epsilon) p_{1,\uparrow} + (R_{32}-\epsilon) p_{2,\uparrow}
-(R_{13}+R_{23}) p_{2,\uparrow} -\gamma p_{3,\uparrow}+ \gamma p_{3,\downarrow},\\
\frac{d p_{3,\downarrow}}{d t}=(R_{13}-\epsilon) p_{1,\downarrow} + (R_{32}+\epsilon) p_{2,\downarrow} 
-(R_{13}+R_{23}) p_{2,\downarrow} -\gamma p_{3,\downarrow}+ \gamma p_{3,\uparrow}.
\end{split}
\end{align}
\end{widetext}
A concise form of the above expressions of the joint occupation probability is given in Eq.~(2). With the joint occupation probabilities, let us now define the quantities: total occupation probability ($P_i$) of being in site $i$ and polarisation of a site ($P^*_i$) as 
\begin{align}
P_i=p_{1,\uparrow}+p_{1,\downarrow},
\end{align}
and
\begin{align}
P^*_i=p_{1,\uparrow}-p_{1,\downarrow},
\end{align}
respectively. With the above definitions of $P_i$ and $P^*_i$, their time evolutions can be written as

\begin{widetext}
\begin{align}
\begin{split}
\frac{d P_{1}}{d t}&=-(R_{21}+R_{31})P_1 + R_{12} P_2 + R_{13} P_3 + \epsilon (P^*_2-P^*_3),\\
\frac{d P_{2}}{d t}&=R_{21}P_1 -( R_{12} + R_{32}) P_2 + R_{23} P_3 + \epsilon (P^*_3-P^*_1),\\
\frac{d P_{3}}{d t}&=R_{31} P_1 + R_{32} P_2 - (R_{13}+R_{23}) P_3 + \epsilon (P^*_1-P^*_2),\\
\frac{d P^*_{1}}{d t}&=-(R_{21}+R_{31})P^*_1 + R_{12} P^*_2 + R_{13} P^*_3 + \epsilon (P_2-P_3) - 2 \gamma P^*_1,\\
\frac{d P^*_{2}}{d t}&=R_{21}P^*_1 -( R_{12} + R_{32}) P^*_2 + R_{23} P^*_3 + \epsilon (P_3-P_1) - 2 \gamma P^*_2,\\
\frac{d P^*_{3}}{d t}&=R_{31} P^*_1 + R_{32} P^*_2 - (R_{13}+R_{23}) P^*_3 + \epsilon (P_1-P_2) - 2 \gamma P^*_3.
\end{split}
\end{align}
\end{widetext}
The above set of equations can be written in a concise matrix form as
\begin{align}
\frac{d \boldsymbol{P} (T,t)}{dt}=\boldsymbol{W}(T_b)  \boldsymbol{P}(T,t), \label{sm: matrix form}
\end{align}
where the vector $\boldsymbol{P}=(P_1, P_2, P_3, P^*_1, P^*_2, P^*_3)^{\mathbb{T}}$ denotes the instantaneous probability distributions of the system at time $t$ and $\boldsymbol{W}(T_b)$ is the transition matrix that is determined at the bath temperature $T_b$, and is given by
\begin{widetext}
\begin{flalign}
\resizebox{\textwidth}{!}{$
\boldsymbol{W}\!= 
\left[\begin{array}{cccccc} 
 -(R_{21}+R_{31})&R_{12} &R_{13} & 0 & \epsilon & -\epsilon  \\
R_{21}& -( R_{12} + R_{32}) &R_{23} & -\epsilon & 0 &\epsilon \\
R_{31} & R_{32} & - (R_{13}+R_{23}) &\epsilon & -\epsilon &0 \\
 0 & \epsilon & -\epsilon& -(R_{21}+R_{31}+2 \gamma)&R_{12} &R_{13}  \\
-\epsilon & 0 &\epsilon&R_{21}& -( R_{12} + R_{32}+2\gamma) &R_{23}  \\
\epsilon & -\epsilon &0& R_{31} & R_{32} & - (R_{13}+R_{23}+2\gamma)  
\end{array}\right].
$}
\label{sm: W matrix}
\end{flalign}
\end{widetext}

\section{\label{sec:simple 3 state without detailed balance}Emergence of complex eigenspectrum}

In this section, we discuss the nature of eigenspectrum of the transition matrix depending on the presence or absence of the detailed balance. It is known that the eigenvalues of a transition matrix are real as long as the rates satisfy detailed balance \cite{schnakenberg1976network}. On the contrary, complex eigenspectrum can emerge in the absence of detailed balance. In the following, we will consider multiple scenarios of the transition matrix: (a) with detailed balance (see Sec.~\ref{supp: with db}) and (b) without detailed balance - with and without the presence of activity (see Sec.~\ref{supp: without db}), to demonstrate different variations of its eigenspectrum.

\subsection{\label{supp: with db}With detailed balance}
\textit{Three state model without activity:} To start with, we first consider the simplest three state model without activity ($\gamma=\epsilon=0$). A walker hops to the energy states with rates $R_{ij}'s$ that satisfy detailed balance, i.e., $R_{ij}P_j=R_{ji}P_i$, where $P_i$ is the occupation probability of the walker in state $i$. As such, these rates take the following form:
\begin{align}
\label{supp: DB}
\begin{split}
R_{ij}=e^{-\frac{B_{ij}-Ej}{k_B T_b}}, \text{ for } i\neq j,\\
R_{ji}=-\sum_{k\neq i}R_{ki}, \text{ for } i= j,
\end{split}
\end{align}
where $E_i$ are the energy levels of the three states and $B_{ij}$  are the barrier heights between any two states $i$ and $j$. Note that $B_{ij}=B_{ji}$ and $R_{ii}$ are chosen as in the above equation such that the normalisation $\sum_i  P_i(t)=1$ is maintained. Although one can in general prove that the eigenvalues are real for a transition matrix that satisfy detailed balance, here we demonstrate it through an example. Let's consider the following model parameters for which the transition matrix satisfy the above mentioned properties: $E_1=0$, $E_2=T_b$, $E_3=3 T_b$, $B_{12}=B_{21}=6 T_b$, $B_{13}=B_{31}=T_b$, $B_{23}=B_{32}=6 T_b$ and $T_b=0.1$. The corresponding eigenvalues of the transition matrix are purely real: $\lambda_1=0,~\lambda_2=-0.0181482$ and $\lambda_3=-7.80453$.

\subsection{\label{supp: without db}Without detailed balance}
\textit{Three state model without activity:} With the simple three state model without activity, let us now consider that the hopping rates $R_{ij}$'s of the walker among the energy states \textit{do not} satisfy detailed balance. It can be achieved by choosing $R_{ij}$'s arbitrarily instead of following Eq.~(\ref{supp: DB}) such that $R_{ij}P_j\neq R_{ji}P_i$. In this case, we show that for certain chosen rates $R_{ij}$'s, the system shows complex eigenspectrum. 

For our example, we consider the following rates: $R_{13}=5 R_{12}$, $R_{21}=100 R_{12}$, $R_{23}=20 R_{12}$, $R_{31}=R_{12}$, $R_{32}=20 R_{12}$, $R_{12}=0.1$ whereas $R_{11}=-R_{21}-R_{31}$, $R_{22}=-R_{12}-R_{32}$ and $R_{33}=-R_{13}-R_{23}$. Here again the rates $R_{ii}$ are chosen as such to maintain the normalization of the occupation probabilities. The eigenvalues obtained for these model parameters are $complex$: $\lambda_1$=0.0, $\lambda_2=-9.6+i 2.99166$ and $\lambda_3=-9.6-i2.99166$. Thus, the above example is sufficient to demonstrate the emergence of complex eigenspectrum in the absence of detailed balance.  In the presence of these complex eigenvalues, the solution of the occupation probabilities starting from the initial conditions $P_1(t=0)=1.0,~P_2(t=0)=0.0~\text{and}~P_3(t=0)=0.0$ are given by:
\bea
\begin{split}
P_1(t)&=0.105825 \Big(1+8.44953~e^{-9.6 t}~  \text{cos}(2.99166 t) \\
&\ - 4.78824~e^{-9.6 t}~ \text{sin}(2.99166 t)\Big), \\
P_2(t)&=0.694293\Big(1-~ e^{-9.6 t}~  \text{cos}(2.99166 t) \\
&\ + 1.60551~ e^{-9.6 t}~ \text{sin}(2.99166 t)\Big), \\
P_3(t)&=0.199881\Big(1-0.199881~ e^{-9.6 t}~  \text{cos}(2.99166 t) \\
&\ -0.607978 ~ e^{-9.6 t}~ \text{sin}(2.99166 t)\Big).
\end{split}
\eea

The presence of the oscillatory terms in the transients of the occupation probabilities also indicates about the existence of oscillations in the transients of the relaxation process as will be discussed later in Sec.~\ref{sec:complex}.

\textit{Three state model with activity:} Let us now consider the three state model with activity ($\epsilon \neq 0$ and $\gamma \neq 0$). The hopping of the walker between the various states will be carried out by the intrinsic rates $R_{ij}'s$ that satisfy detailed balance following Eq.~(\ref{supp: DB}), and also due to the active rates $\epsilon$ and $\gamma$. The effective rate of hopping between the various states is now governed by the transition matrix $\boldsymbol{W}$ [see Eq.~(\ref{sm: matrix form})] whose elements are denoted by $W_{ij}$. Thus, the effective transition rates, due to the active component, are not pairwise balanced between the states. As a result, the effective rate $\boldsymbol{W}$  does not satisfy detailed balance, i.e., $W_{ij}P_j\neq W_{ji}P_i$ holds true for this case and as such the eigenspectrum of $\boldsymbol{W}$ can be complex. 


Let's consider the following model parameters for which the transition matrix $\boldsymbol{W}$ satisfy the above mentioned properties: $E_1=0$, $E_2=T_b$, $E_3=3 T_b$, $B_{12}=B_{21}=4 T_b$, $B_{13}=B_{31}=5 T_b$, $B_{23}=B_{32}=4 T_b$ and $T_b=0.1$. The corresponding eigenspectrum of the transition matrix is complex: $\lambda_1=0$, $\lambda_2=-2.31392 - i 2.63085$, $\lambda_3=-2.31392 +i 2.63085$, $\lambda_4=-2.31392 -i 2.99528$, $\lambda_5=-2.31392 +i 2.99528$ and $\lambda_6=-4.0$. Similar to the previous case where complex eigenvalues lead to the presence of oscillations in the transient of the relaxation process, here also one obtains oscillatory terms in the solution of $\boldsymbol{P}(t)$ due to the presence of complex eigenspectrum and hence  will eventually lead to oscillations in the transients as will be discussed later in Sec.~\ref{sec:complex}.

\section{\label{sec:a2}Calculation of the coefficient $a_2(T,T_b)$}

Unless the eigenvalues of a system's transition matrix are \textit{degenerate} or \textit{complex}, its relaxation is governed by the second largest eigenvalue: $\text{Re}\lambda_2$, as the contribution of all other terms in the expansion of Eq.~(\ref{time evolution}) will be exponentially suppressed. In particular, the existence of the Mpemba effect can be determined by the coefficient $a_2(T, T_b)$ in Eq.~(\ref{time evolution}) associated with the second largest eigenvalue. But before calculating the coefficient 
 $a_2(T, T_b)$, let us first perform a similarity transformation for the matrix $\boldsymbol{W}$ that transforms it to a symmetric matrix $\boldsymbol{\tilde{W}}$, i.e.,
\begin{equation}
\boldsymbol{\tilde{W}}=\boldsymbol{X}^{-1} \boldsymbol{W} \boldsymbol{X},
\end{equation}
where the matrix $\boldsymbol{X}$ is formed from the eigenvectors of $\boldsymbol{W}$. Note that the symmetric matrix $\boldsymbol{\tilde{W}}$ has the same eigenvalues as $\boldsymbol{W}$ but now it has orthogonal set of real eigenvectors denoted by $\boldsymbol{v}'$. The eigenvectors $\boldsymbol{v}'$ of matrix $\boldsymbol{\tilde{W}}$ are obtained from the transformation:
\begin{equation}
\boldsymbol{v}'=\boldsymbol{X}^{-1} \boldsymbol{v}, \label{eigenvector transform}
\end{equation}
and they satisfy the orthogonality relation
\begin{equation}
\boldsymbol{v}'_i. \boldsymbol{v}'_j=\delta_{ij}.
\end{equation}

The transformed eigenvectors $\boldsymbol{v}'$ obtained using Eq.~(\ref{eigenvector transform}) will be useful in the analysis of the Mpemba effect. Since the relaxation is dominated by the second largest eigenvalue, we can use the following approximation for Eq.~(\ref{time evolution})
\begin{equation}
\boldsymbol{P}(T, t)\simeq \boldsymbol{\pi}(T_b)+a_2(T, T_b) \boldsymbol{v}_2 e^{\lambda_2 t}. \label{approx time evolution}
\end{equation}

Now, let us consider two identical systems which are prepared at two different initial steady states corresponding to the choice of $hot$ and $cold$ bath temperatures $T_h$ and $T_c$ respectively. As both the systems are quenched to the common steady state of the bath temperature $T_b$ such that $T_h>T_c>T_b$ then the Mpemba effect is said to exist if
\begin{equation}
|a_2(T_h,T_b)|<|a_2(T_c,T_b)|. \label{condition}
\end{equation}
It is so because with the above condition, the distribution $\boldsymbol{P}(T_c, t)$ of the initially cold system essentially lags behind the distribution $\boldsymbol{P}(T_h, t)$ of the initially hot system leading to the Mpemba like relaxation. Thus, the overall task reduces to computing the coefficient $a_2(T, T_b)$ in Eq.~(\ref{approx time evolution}). Now, in order to derive  $a_2(T,T_b)$, we multiply Eq.~(\ref{approx time evolution}) with $\boldsymbol{v}'_2. \boldsymbol{X}^{-1}$ at time $t=0$:
\begin{align}
&\boldsymbol{v}'_2. \boldsymbol{X}^{-1} \boldsymbol{P}(T, t=0) \nonumber\\
&= \boldsymbol{v}'_2. \boldsymbol{X}^{-1} \boldsymbol{\pi} (T_b) + a_2(T,T_b)  \boldsymbol{v}'_2. \boldsymbol{X}^{-1}  \boldsymbol{v}_2, \nonumber \\
&=\boldsymbol{v}'_2.\boldsymbol{v}'_1 + a_2(T, T_b) \boldsymbol{v}'_2.\boldsymbol{v}'_2, \label{a2 calculation}
\end{align}
where we have used the relations: $\boldsymbol{X}^{-1} \boldsymbol{\pi} (T_b)=\boldsymbol{v}'_1$ and $\boldsymbol{X}^{-1}  \boldsymbol{v}_2=\boldsymbol{v}'_2$. Since the eigenvectors $\boldsymbol{v}'_i$ are orthogonal by construction, we can write \eref{a2 calculation} as:
\be
a_2(T,T_b)=\frac{\boldsymbol{v}'_2. \boldsymbol{X}^{-1} \boldsymbol{P}(T,t=0)}{||\boldsymbol{v}'_2||^2}, \label{a2}
\ee
where we have defined $||\boldsymbol{v}'_2||^2=\boldsymbol{v}'_2.\boldsymbol{v}'_2$. Thus having obtained the expression for $a_2(T,T_b)$ in \eref{a2}, we now have to look for the non-monotonicity in the behaviour of $a_2(T,T_b)$ with $T$ according to \eref{condition} in order to quantify the existence of the Mpemba effect. However, we will show that the above criterion of the Mpemba effect is only useful when the eigenspectrum is real. In order to check for the validity of the above discussion, we will consider the cases of \textit{real} as well as \textit{complex} eigenspectrums, and check for the existence of the Mpemba effect in terms of the criteria of $a_2(T, T_b)$ as well as in terms of other distance measures.

\section{\label{sec:real}Mpemba effect in terms of $a_2$ criteria}
In this section, we consider two different cases where the transition matrix can have: (a) real and (b) complex eigenspectrum. In both the cases, we check if the inference of the Mpemba effect in terms of $a_2$-criteria is consistent with that obtained from other distance measures such as $L_1$-norm or KL-divergence. 
\subsection{Case of real eigensystem: $a_2$ criteria is sufficient}
To start with, we first consider a case where the eigenvalues of the transition matrix $\boldsymbol{W}$ are \textit{real} and \textit{non-degenerate}. It is in fact always true for equilibrium systems~\cite{schnakenberg1976network}. In such a scenario, the criteria for the existence of the Mpemba effect is described solely in terms of the behavior of $|a_2(T,T_b)|$ as described in the previous section. As we find out, the requirement of having real eigensystem in the present non-equilibrium model of active Markov chains, is satisfied by the following choice of the model parameters: $E_1=0,~E_2=T_b,~E_3=3 T_b,~B_{12}=B_{21}=4 T_b,~B_{13}=B_{31}=7 T_b$, and $B_{32}=B_{23}=4 T_b$. The activity parameters characterising the active random walker are chosen to be $\gamma/R_{12}(T_b)=60.27$ and $\epsilon/R_{12}(T_b)=20.09$ while the bath temperature corresponding to the final quenched state is $T_b=0.1$.

\begin{figure}
\centering
\includegraphics[width=\columnwidth]{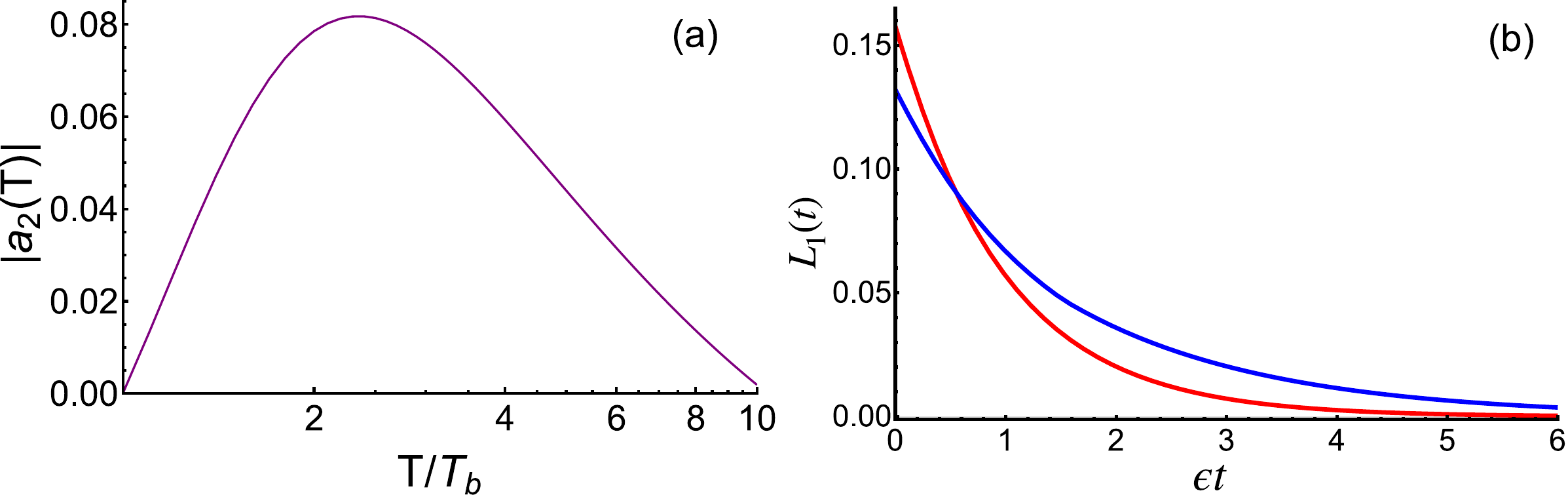}
\caption{(a) Non-monotonic variation of $|a_2(T,T_b)|$ with temperature $T$ for the active Markov chain illustrates the presence of the Mpemba effect. (b) It is also confirmed in terms of the crossing of $L_1$-norm as distance function for a choice of two initial temperatures from the plot (a) where $|a_2(T,T_b)|$ is non-monotonous with $T$. The rates $R_{ij}$ are determined by $E_1=0,~E_2=T_b,~E_3=3 T_b,~B_{12}=B_{21}=4 T_b,~B_{13}=B_{31}=7 T_b$, and $B_{32}=B_{23}=4 T_b$. The activity parameters characterising the active random walker are chosen to be $\gamma/R_{12}(T_b)=60.27$ and $\epsilon/R_{12}(T_b)=20.09$. The bath temperature corresponding to the final quenched state is $T_b=0.1$.} \label{supp: active mpemba}
\end{figure}

For this case, the coefficient $|a_2(T,T_b)|$ decays non-monotonically with temperature as shown in Fig.~\ref{supp: active mpemba}(a) and thus exhibits the signature of the Mpemba effect. Now having shown the non-monotonic behavior of $|a_2(T, T_b)|$, we show that the criteria of identifying the Mpemba effect in terms of $|a_2(T,T_b)|$ is sufficient and the same identification is reflected in terms of the crossing of distance function as well, as shown in Fig.~\ref{supp: active mpemba} (b) with $L_1$-norm.

\begin{figure}
\centering
\includegraphics[width=\columnwidth]{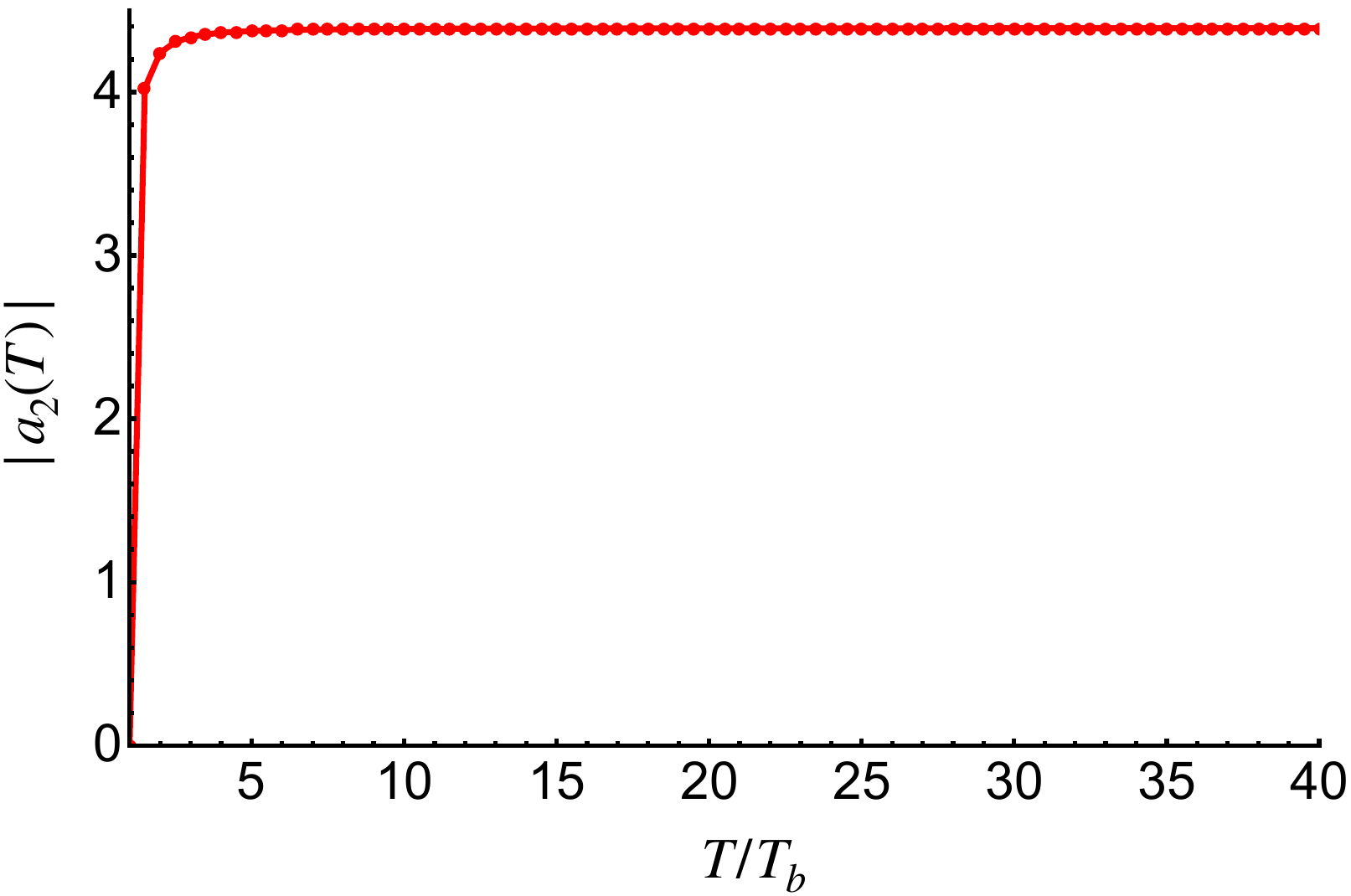}
\caption{Monotonic increase of $|a_2(T,T_b)|$ with temperature $T$ for the system of active Markov chain indicates the absence of the Mpemba effect. Here, the model parameters are:  $E_1=0,~E_2=T_b,~E_3=3 T_b,~B_{12}=B_{21}=6 T_b,~B_{13}=B_{31}= T_b$, and $B_{32}=B_{23}=6 T_b$, $\gamma/R_{12}(T_b)=148.41$, $\epsilon/R_{12}(T_b)=148.41$ and $T_b=0.1$. Although the variation in  $|a_2(T,T_b)|$ for the chosen model parameters do not characterize the double crossing as observed in terms of distance measures [see Fig.~\ref{non-degenerate case multiple crossing}] but correctly determines the absence of the Mpemba effect.} \label{supp: a2 real eigensystem multiple crossing}
\end{figure}

Next we consider the case where the eigenspectrum is again real and non-degenerate. However, we observe a double crossing in this case in the time evolution of the $L_1$-norm  of the initially $hot$ and the $cold$ system as illustrated in Fig.~\ref{non-degenerate case multiple crossing}. Such a scenario depicts the absence of the Mpemba effect at long times. The same behavior is captured by the $|a_2(T,T_b)|$ criteria which also shows a monotonic behavior with $T$ as illustrated in Fig.~\ref{supp: a2 real eigensystem multiple crossing}, and hence correctly predicts the absence of the Mpemba effect in this case. Thus, we can infer that there is an one to one correspondence between the conclusions drawn from $a_2$ criteria and any other distance measure about the overall existence of the Mpemba effect in the system of active Markov chain with \textit{real eigenspectrum}.

\subsection{\label{subsec:a2 vs gamma and epsilon}$a_2$ as a function of the active rates $\epsilon$ and $\gamma$}
\begin{figure}
\centering
\includegraphics[width=\columnwidth]{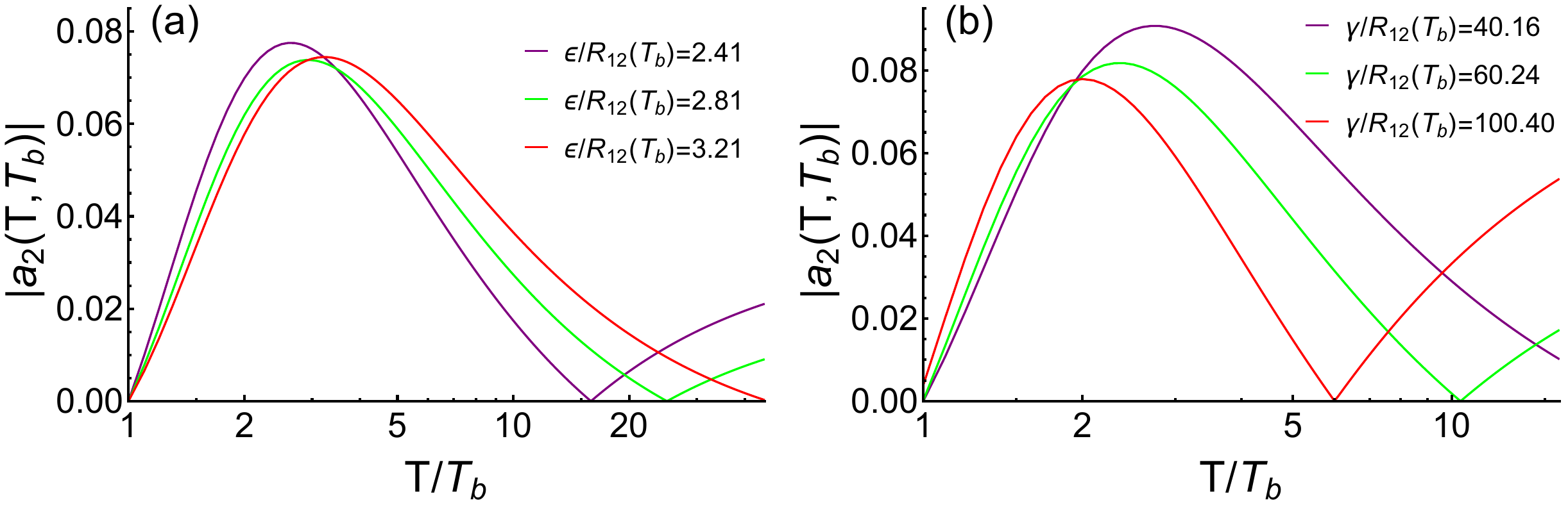}
\caption{\textbf{Analysis of the variation of $|a_2(T,T_b)|$ with temperature $T$ as the active rates $\epsilon$ and $\gamma$ are varied}. Panel (a) illustrates the variation of $|a_2(T,T_b)|$ with $T$ for different persistence rates, $\epsilon$ of the active random walker while keeping the flipping rate $\gamma$ fixed at $\gamma/R_{12}(T_b)=60.27$, whereas panel (b) illustrates the variation of $|a_2(T,T_b)|$ with $T$ for different flipping rates, $\gamma$ but keeping $\epsilon/R_{12}(T_b)=20.09$ fixed. The analysis is done in the regime of phase space of system parameters where the eigenspectrum is real corresponding to a particular choice of the parameters for both (a) and (b) as $E_1=0,~E_2=T_b,~E_3=3 T_b,~B_{12}=B_{21}=4 T_b,~B_{13}=B_{31}=7 T_b$, $B_{32}=B_{23}=4 T_b$ and $T_b=0.1$. } \label{supp: variation of a2 with epsilon and gamma}
\end{figure}

In this section, now we explore how the quantity $|a_2(T, T_b)|$ or in other words, the range of initial conditions that lead to  the Mpemba effect is affected by the variation of the active rates $\epsilon$ and $\gamma$. Note that we still confine ourselves to the parameter space where the eigenspectrum of the transition matrix is real. Note that the Mpemba effect occurs for the set of initial temperatures where there is non-monotonic change of $|a_2(T)|$ with $T$. With that, we find that the region of the set of initial temperatures $T$ where such a behavior is observed for $|a_2(T)|$ increases with the increase in the parameter $\epsilon$ [see Fig.~\ref{supp: variation of a2 with epsilon and gamma}(a)]. On the contrary, the behavior of $|a_2(T,T_b)|$ is the opposite with the variation in $\gamma$ [see Fig.~\ref{supp: variation of a2 with epsilon and gamma}(b)] where the range of initial temperatures that show the Mpemba effect decreases with increasing $\gamma$.

\subsection{\label{sec:complex}Case of complex eigensystem \& Oscillatory Mpemba effect: $a_2$ criteria NOT sufficient }

\begin{figure}
\centering
\includegraphics[width=\columnwidth]{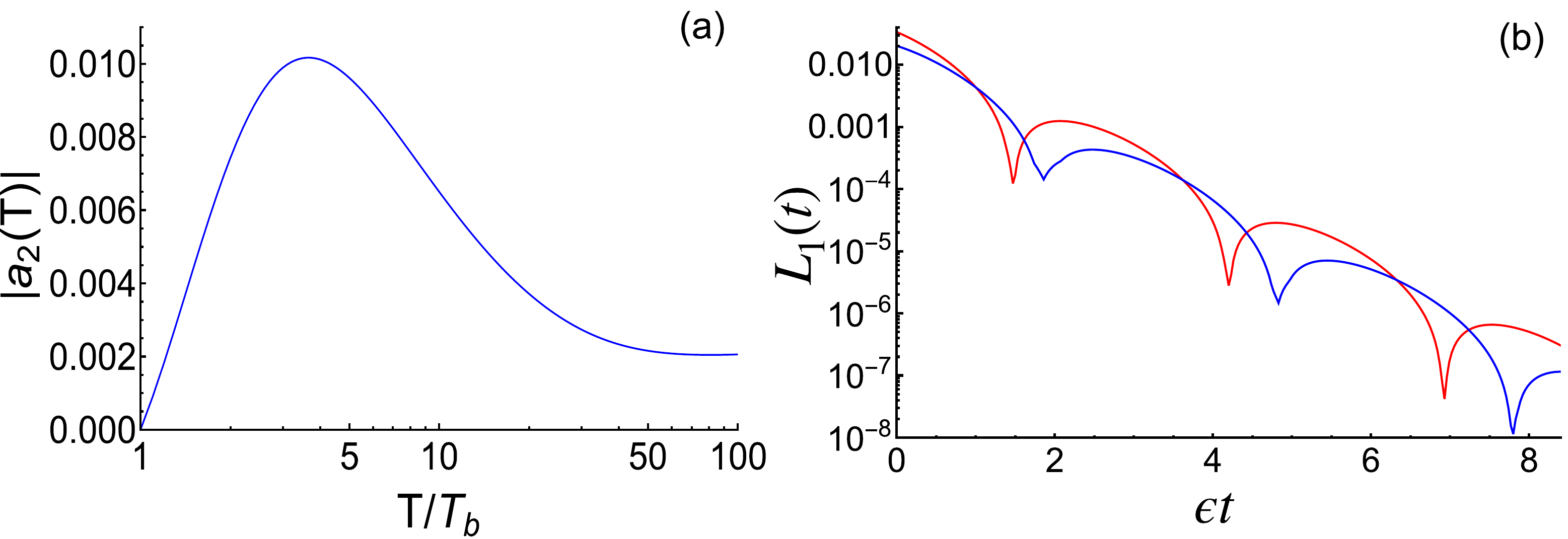}
\caption{\textbf{$a_2$-criteria does not characterize oscillatory Mpemba effect:} (a) Non-monotonic variation of $|a_2(T, T_b)|$ with temperature $T$ for the active Markov chain indicates the presence of the Mpemba effect. (b) Oscillatory behavior in the time evolution of $L_1$-norm is observed  and indicates the ambiguity in the presence of a net Mpemba effect. The model parameters are: $E_1=0,~E_2=T_b,~E_3=3 T_b,~B_{12}=B_{21}=4 T_b,~B_{13}=B_{31}=5 T_b$, and $B_{32}=B_{23}=4 T_b$, $\gamma/R_{12}(T_b)=160.68$, $\epsilon/R_{12}(T_b)=120.51$ and $T_b=0.1$. The temperatures of the initially hot and the cold systems in (b) are chosen to be $T_h/T_b=50$ and $T_c/T_b=5$ respectively where $|a_2(T, T_b)|$ is non-monotonous with $T/T_b$.} \label{supp: active mpemba oscillatory}
\end{figure}

Now, we consider the case where the eigenspectrum  of the transition matrix $\boldsymbol{W}$ can include complex eigenvalues. It corresponds to making an appropriate choice of the model parameters in terms of the active rates: $\epsilon$ and $\gamma$. The presence of complex eigenspectrum leads to the presence of oscillations or multiple crossings of relaxation trajectory across the final steady state before it finally asymptotically reaches the state [see Fig.~\ref{supp: active mpemba oscillatory}(b)]. In that case, we show that the $|a_2(T,T_b)|$ criteria fails to characterize such oscillatory Mpemba effect and one needs to use an appropriate distance measure (e.g., $L_1$-norm) in order to characterize such oscillatory behavior in the Mpemba effect as shown in Fig.~\ref{supp: active mpemba oscillatory}.

%

\end{document}